\pdfoutput=1  
\documentclass[sigconf,natbib,screen]{acmart}
\usepackage[utf8]{inputenc}

\AtBeginDocument{%
  }

\usepackage{booktabs,enumitem,multirow}
\usepackage[normalem]{ulem}
\usepackage{framed,fancybox,ascmac}
\usepackage{wrapfig}
\usepackage{enumitem}
\usepackage{diagbox}

\usepackage{nicefrac}

\usepackage{mathtools}
\usepackage{apxproof}

\usepackage{comment}
\usepackage{url}
\usepackage[caption=false]{subfig}

\usepackage{xspace}

\usepackage{bm}

\usepackage[nameinlink]{cleveref}
\usepackage{nameref}

\usepackage{tcolorbox}

\usepackage{lineno}

\newtheorem{theorem}{Theorem}[section]
\newtheorem*{theorem*}{Theorem}

\newtheorem{claim}[theorem]{Claim}

\theoremstyle{definition}
\newtheorem{definition}[theorem]{Definition}

\crefname{theorem}{Theorem}{Theorems}
\crefname{lemma}{Lemma}{Lemmas}
\crefname{claim}{Claim}{Claims}
\crefname{remark}{Remark}{Remarks}
\crefname{observation}{Observation}{Observations}
\crefname{corollary}{Corollary}{Corollaries}
\crefname{appendix}{Appendix}{Appendices}
\crefname{section}{Section}{Sections}
\crefname{algorithm}{Algorithm}{Algorithms}
\crefname{equation}{Eq.}{Eqs.}
\crefname{figure}{Figure}{Figures}
\crefname{table}{Table}{Tables}

\crefformat{footnote}{#2\textsuperscript{\normalfont #1}#3}

\usepackage{algorithmicx}
\usepackage[ruled]{algorithm}
\usepackage[noend]{algpseudocode}
\algnewcommand\And{\; \textbf{and} \;}
\algnewcommand\Or{\; \textbf{or} \;}
\algnewcommand\To{\; \textbf{to} \;}
\algnewcommand\Continue{\textbf{continue}}
\algnewcommand\Not{\textbf{not}}

\DeclareMathOperator*{\argmax}{argmax}
\DeclareMathOperator*{\median}{median}
\newcommand{\med}{\mathrm{med}}
\renewcommand{\vec}[1]{\mathbf{\bm{#1}}}

\renewcommand{\vec}[1]{\mathbf{\bm{#1}}}
\newcommand{\mat}[1]{\mathbf{\bm{#1}}}
\newcommand{\bigO}{\mathcal{O}}

\newcommand{\ILD}{\textsf{\textup{ILD}}\xspace}
\newcommand{\disp}{\textsf{\textup{disp}}\xspace}
\newcommand{\GILD}{\textsf{\textup{GILD}}\xspace}
\newcommand{\Random}{\textsf{\textup{Random}}\xspace}

\newcommand{\nDCG}{\textsf{\textup{nDCG}}\xspace}
\newcommand{\nILD}{\textsf{\textup{nILD}}\xspace}
\newcommand{\ndisp}{\textsf{\textup{ndisp}}\xspace}

\newcommand{\OPT}{{\mathrm{OPT}}}
\newcommand{\Gr}{{\mathrm{Gr}}}
\DeclareMathOperator{\divf}{\mathop{\mathsf{f}}}
\DeclareMathOperator{\divg}{\mathop{\mathsf{g}}}

\newcommand{\ML}{\textsl{ML-1M}\xspace}
\newcommand{\Ama}{\textsl{Amazon}\xspace}

\newcommand{\circles}{\textsl{TwoCircles}\xspace}
\newcommand{\ellipse}{\textsl{Ellipse}\xspace}
\newcommand{\feedback}{\textsl{feedback}\xspace}
\newcommand{\genre}{\textsl{genre}\xspace}
\newcommand{\easer}{$\textsc{ease}^\textsc{r}$\xspace}

\newcommand{\bbR}{\mathbb{R}}

\usepackage{balance}
\usepackage{multicol}

\copyrightyear{2023}
\acmYear{2023}
\setcopyright{rightsretained}
\acmConference[SIGIR '23]{Proceedings of the 46th International ACM SIGIR Conference on Research and Development in Information Retrieval}{July 23--27, 2023}{Taipei, Taiwan}
\acmBooktitle{Proceedings of the 46th International ACM SIGIR Conference on Research and Development in Information Retrieval (SIGIR '23), July 23--27, 2023, Taipei, Taiwan}\acmDOI{10.1145/3539618.3591623}
\acmISBN{978-1-4503-9408-6/23/07}



\makeatletter
\gdef\@copyrightpermission{
  \begin{minipage}{0.3\columnwidth}
   \href{https://creativecommons.org/licenses/by/4.0/}{\includegraphics[width=0.90\textwidth]{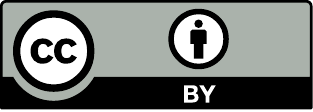}}  
  \end{minipage}\hfill
  \begin{minipage}{0.7\columnwidth}
   \href{https://creativecommons.org/licenses/by/4.0/}{This work is licensed under a Creative Commons Attribution International 4.0 License.}
  \end{minipage}
  \vspace{5pt}
}
\makeatother

\begin{document}

\title{A Critical Reexamination of Intra-List Distance and Dispersion}
\author{Naoto Ohsaka}
\affiliation{
    \institution{CyberAgent, Inc.}
    \city{Tokyo}
    \country{Japan}
}
\email{ohsaka_naoto@cyberagent.co.jp}
\orcid{0000-0001-9584-4764}

\author{Riku Togashi}
\affiliation{
    \institution{CyberAgent, Inc.}
    \city{Tokyo}
    \country{Japan}
}
\email{rtogashi@acm.org}
\orcid{0000-0001-9026-0495}

\begin{abstract}
Diversification of recommendation results is a promising approach for coping with the uncertainty associated with users' information needs.
Of particular importance in diversified recommendation is to define and optimize an appropriate diversity objective.
In this study, we revisit the most popular diversity objective
called \emph{intra-list distance (ILD)}, defined as
the average pairwise distance between selected items, and
a similar but lesser known objective called \emph{dispersion},
which is the minimum pairwise distance.
Owing to their simplicity and flexibility,
ILD and dispersion have been used in a plethora of diversified recommendation research.
Nevertheless, \emph{we do not actually know} what kind of items are preferred by them.

We present a critical reexamination of ILD and dispersion from theoretical and experimental perspectives.
Our theoretical results reveal that
these objectives have potential drawbacks:
ILD may select \emph{duplicate} items that are very close to each other,
whereas dispersion may overlook \emph{distant} item pairs.
As a competitor to ILD and dispersion,
we design a diversity objective called Gaussian ILD,
which can \emph{interpolate} between ILD and dispersion by tuning the bandwidth parameter.
We verify our theoretical results by experimental results using real-world data
and confirm the extreme behavior of ILD and dispersion in practice.
\end{abstract}

\maketitle

\section{Introduction}
\label{sec:intro}

In recommender systems, solely improving the prediction accuracy of user preferences,
as a single objective, is known to have
the risk of recommending over-specialized items to a user,
resulting in low user satisfaction \cite{mcnee2006being}.
The primary approach for addressing such issues
arising from the uncertainty associated with users' information needs
is the introduction of \emph{beyond-accuracy objectives} \cite{kaminskas2017diversity}
such as diversity, novelty, and serendipity.
Among the most important beyond-accuracy objectives is \emph{diversity},
which refers to the internal differences between items recommended to a user.
Recommending a set of diverse items may increase the chance of satisfying a user's needs.
However, defining diversity is a nontrivial task because
the contribution of a particular item depends on the other selected items.
Of particular importance in diversified recommendation is thus to define and optimize an appropriate diversity objective.

In this study, we revisit two diversity objectives.
One is the \emph{intra-list distance (ILD)},
which is arguably the most frequently used objective for diversity.
The ILD \cite{smyth2001similarity,ziegler2005improving}
is defined as the average pairwise distance between selected items for a particular distance metric.
ILD is easy to use and popular in diversified recommendation research for the following reasons:

\begin{itemize}[leftmargin=*]
\item[\textbf{1.}] It is a \emph{distance-based objective} \cite{drosou2017diversity},
which only requires a pairwise distance metric between items;
thus, we can flexibly adopt any metric depending on the application, e.g., 
the Jaccard distance \cite{yu2009it,gollapudi2009axiomatic,kaminskas2017diversity},
taxonomy-based metric \cite{ziegler2005improving}, and
cosine distance~\cite{cheng2017learning}.
\item[\textbf{2.}] The definition is ``intuitive'' in that
it simply integrates pairwise distances between items in a recommendation result.
\item[\textbf{3.}] Although maximization of ILD is \textbf{NP}-hard \cite{tamir1991obnoxious},
a simple greedy heuristic efficiently identifies an item set with a nearly optimal ILD \cite{ravi1994heuristic,birnbaum2009improved}.
This heuristic can be easily incorporated into recommendation algorithms
\cite{yu2009it,gollapudi2009axiomatic,hurley2011novelty,wasilewski2016incorporating,sha2016framework,cheng2017learning}.
\end{itemize}
Indeed, ILD appears in many surveys on diversified recommendations \cite{kaminskas2017diversity,wu2019recent,castells2015novelty,drosou2010search,drosou2017diversity,kunaver2017diversity}.
The other objective investigated in this study is 
a similar but lesser known one called \emph{dispersion},
which is defined as the \emph{minimum} pairwise distance between selected items.
Although dispersion seldom appears in the recommendation literature \cite{gollapudi2009axiomatic,drosou2010search},
it has the aforementioned advantages.
Nevertheless, \emph{we do not actually know}
what kind of items are preferred by ILD and dispersion; for instance:
Are the items selected by optimizing ILD or dispersion satisfactorily distant from each other?
What if the entire item set is clustered or dispersed?

\begin{figure}[tbp]
    \centering
    \includegraphics[width=0.65\hsize]{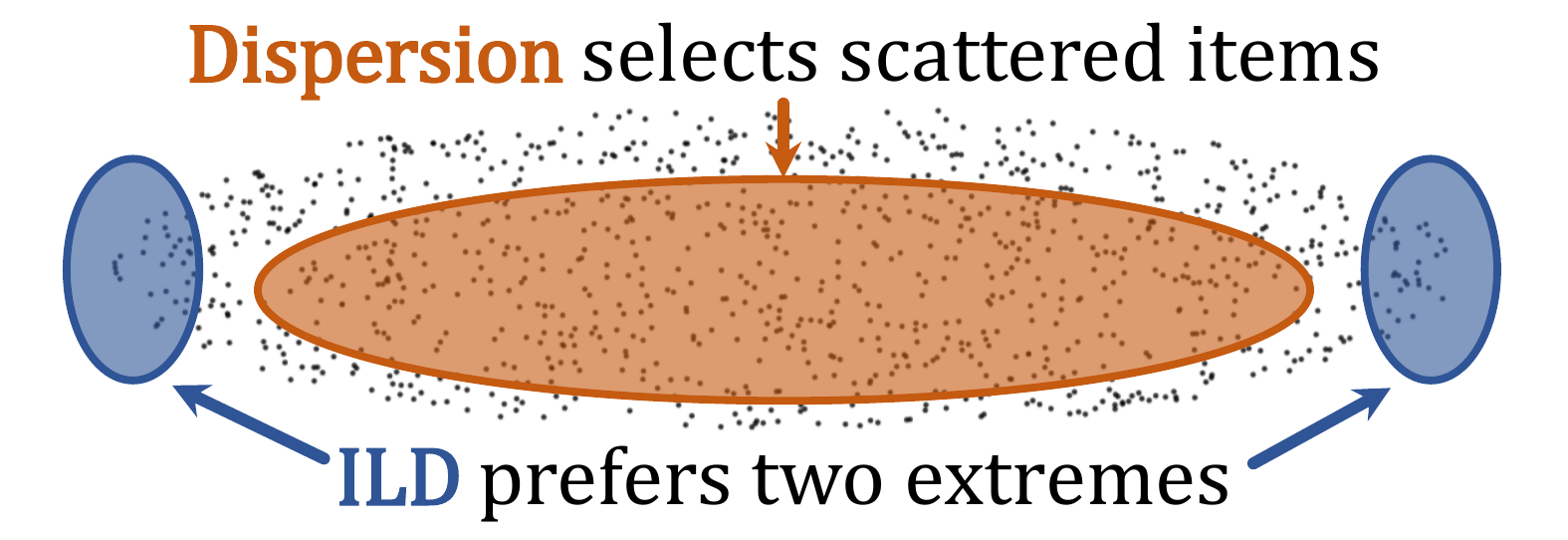}
    \caption{An example such that intra-list distance and dispersion select very different items.}
    \label{fig:intro:lesson}
\end{figure}

\subsection{Our Contributions}
This study presents a critical reexamination of ILD and dispersion
from both theoretical and experimental perspectives.
To answer the aforementioned questions,
we investigate whether enhancing one (e.g., ILD) leads to an increase in the other (e.g., dispersion),
in the hope that
we can characterize what they are representing and reveal their drawbacks.
We first identify the following potential drawbacks of ILD and dispersion based on our theoretical comparisons (\cref{sec:theory}):
\emph{ILD selects items in a well-balanced manner if
the entire item set is separated into two clusters.
However, it may generally select duplicate items that are very close to each other.
The items chosen by dispersion are well-scattered, but distant item pairs may be overlooked.}

We then conduct numerical experiments to verify
the assertions based on our theoretical analysis (\cref{sec:practice}).
Our empirical results using MovieLens~\cite{harper2015movielens} and
Amazon Review \cite{ni2019justifying} demonstrate that
\emph{ILD can readily select many items that are similar or even identical},
which is undesirable if we wish to recommend very few items.
\cref{fig:intro:lesson} shows a cloud of points in an ellipse such that
ILD and dispersion select very different item sets.
Our theoretical and empirical results imply that
the items selected via ILD are biased toward two distant groups;
items in the middle of the ellipse are never chosen.
In contrast,
the items selected by dispersion are well-scattered.

To better understand the empirical behaviors of ILD and dispersion,
we design a new distance-based objective that generalizes ILD and dispersion
as a competitor (\cref{sec:gauss}).
The designed one, \emph{Gaussian ILD (GILD)},
is defined as the average of the Gaussian kernel distances \cite{phillips2011gentle} between selected items.
GILD has bandwidth parameter $\sigma$, and
we prove that GILD approaches ILD as $\sigma \to \infty$ and
approaches dispersion as $\sigma \to 0$; i.e.,
it can \emph{interpolate} between them.
We experimentally confirm that
GILD partially circumvents the issues caused by the extreme behaviors of ILD and dispersion,
thereby achieving a \emph{sweet spot} between them (\cref{sec:practice}).

Finally, we examine the recommendation results obtained by enhancing ILD, dispersion, and GILD (\cref{sec:recommend}).
The experimental results demonstrate that 
(1) ILD frequently selects duplicate items, and thus it is not an appropriate choice;
(2) if the relevance of the recommended items is highly prioritized,
dispersion fails to diversify the recommendation results for some users.

In summary,
\emph{ILD is not appropriate for either evaluating or enhancing distance-based diversity}, whereas
\emph{dispersion is often suitable for improving diversity, but not necessarily for measuring diversity}.

\section{Related Work}\label{sec:related}

Diversity enhancement has various motivations \cite{castells2015novelty};
e.g.,
(1)
because a user's preference is uncertain owing to the inherent sparsity of user feedback,
recommending a set of diverse items has the potential to satisfy a user's needs;
(2)
users desire diversity of recommended items due to the variety-seeking behavior.
Other beyond-accuracy objectives include novelty, serendipity, and coverage;
see, e.g., \citet*{castells2015novelty}, \citet*{kaminskas2017diversity}, and
\citet*{zangerle2022evaluating}.

Generally, there are two types of diversity.
One is \emph{individual diversity}, which represents the diversity of recommended items for each user.
The other is \emph{aggregate diversity} 
\cite{adomavicius2012improving,adomavicius2014optimization},
which represents the diversity \emph{across} users and promotes long-tail items.
We review the definitions and enhancement algorithms for individual diversity,
which is simply referred to as \emph{diversity} throughout this paper.

\paragraph{Defining Diversity Objectives}
The \emph{intra-list distance} (ILD)
(also known as the \emph{average pairwise distance}) 
due to \citet*{smyth2001similarity} and \citet*{ziegler2005improving}
is among the earliest diversity objectives in recommendation research.
Owing to its simplicity and flexibility in the choice of a distance metric,
ILD has been used in a plethora of subsequent works \cite{zhang2008avoiding,yu2009it,gollapudi2009axiomatic,hurley2011novelty,boim2011diversification,vargas2011rank,borodin2012max,su2013set,sha2016framework,cheng2017learning,ekstrand2014user}.
\emph{Dispersion} is another distance-based diversity objective that is similar to ILD.
Maximizing the dispersion value is known as
the \emph{$p$-dispersion} problem in operations research
and is motivated by applications in facility location 
\cite{kuby1987programming,erkut1989analytical,erkut1990discrete,ravi1994heuristic}.
Notably,
only a few studies on recommender systems
\cite{gollapudi2009axiomatic,drosou2010search} adopt dispersion as the diversity objective.
\emph{Determinantal point processes (DPP)} are probabilistic models that express
the negative correlation among items using the determinant \cite{macchi1975coincidence,borodin2005eynard}.
DPP-based objectives have recently been applied to recommender systems \cite{qin2013promoting}.
See \citet*{kulesza2012determinantal} for more details.
\emph{Topical diversity objectives} use predefined topic information to directly evaluate
how many topics are covered by selected items and/or
the extent to which topic redundancy should be avoided
\cite{agrawal2009diversifying,vargas2014coverage,ashkan2015optimal,antikacioglu2019new}.
Such topic information is often readily available
in many domains such as movies, music, and books.
In this paper,
we do not compare DPPs or topical diversity
because we deeply investigate ILD and dispersion, which are more commonly used.

\citet{gollapudi2009axiomatic} use an \emph{axiomatic approach}, in which
they design a set of axioms that
a diversity objective should satisfy,
and
prove that any objective,
including ILD and dispersion, cannot satisfy all the axioms simultaneously.
\citet{amigo2018axiomatic} present another axiomatic analysis of diversity-aware evaluation measures.
Our study is orthogonal to these works because
we focus on elucidating what diversity objectives represent.

\paragraph{Diversity Enhancement Algorithms}
We review algorithms for enhancing the diversity of recommended items.
The basic approach simultaneously optimizes both relevance and diversity.
Given the relevance $\mathsf{rel}(i)$ for each item $i$ and 
a diversity objective $\mathsf{div}(\cdot)$ (e.g., ILD),
we can formulate an objective function as
a linear combination of the average relevance and diversity of selected items $S$, i.e.,
\begin{align}
\label{eq:rel-div}
\max_{S} \; (1-\lambda) \cdot \frac{1}{|S|}\sum_{i \in S}\mathsf{rel}(i) + \lambda \cdot \mathsf{div}(S),
\end{align}
where $\lambda \in (0,1)$ is the trade-off parameter.
The \emph{maximal marginal relevance (MMR)} \cite{carbonell1998use} is an initial attempt using this approach,
which applies a greedy heuristic to \cref{eq:rel-div}.
Greedy-style algorithms are widely used in many diversified recommendation studies \cite{agrawal2009diversifying,vargas2014coverage,ashkan2015optimal,yu2009it,gollapudi2009axiomatic,hurley2011novelty,wasilewski2016incorporating,sha2016framework,cheng2017learning}.
Other algorithms include local search \cite{yu2009it},
binary quadratic programming \cite{zhang2008avoiding,hurley2011novelty}, and
multi-objective optimization \cite{ribeiro2012pareto,ribeiro2014multiobjective}.
However, even (Pareto) optimal solutions are undesirable
unless we choose an ``appropriate'' objective to be optimized.
We investigate whether the greedy maximization of one diversity objective is useful for enhancing another objective.

\emph{Learning-to-rank} approaches aim to directly learn
the optimal ranking of recommended items for each user
under a particular definition of the loss function.
Notably,
the underlying function that models diversity often originates from existing diversity objectives, including
ILD \cite{wasilewski2016incorporating,cheng2017learning}.
Thus, our study helps understand the impact of underlying diversity modeling on recommendation results.

\paragraph{Evaluation Measures in Information Retrieval}
In information retrieval (IR),
efforts were made to render classical IR evaluation measures diversity-aware to
address the uncertainty in users' queries, e.g., 
$\alpha$-normalized discounted cumulative gain ($\alpha$-nDCG)~\cite{clarke2008novelty},
Intent-Aware measures~\cite{agrawal2009diversifying},
D$\sharp$-measures~\cite{sakai2011evaluating}, and
$\alpha\beta$-nDCG~\cite{parapar2021towards}.
We do not consider such diversity-aware IR measures, which assume
that a distribution over the intents is available for each query.

\section{Preliminaries}\label{sec:pre}

\paragraph{Notations}
For a nonnegative integer $n$, let $[n] \triangleq \{1,2,\ldots, n\}$.
For a finite set $S$ and an integer $k$, we write ${S \choose k}$ for the family of all size-$k$ subsets of $S$.
Vectors and matrices are written in bold (e.g., $\vec{v}$ and $\mat{A}$), and
the $i$-th entry of a vector $\vec{v}$ in $\bbR^d$ is denoted $v(i)$.
The Euclidean norm is denoted $\| \cdot \|$;
i.e., $\|\vec{v}\| \triangleq \sqrt{\sum_{i \in [d]}v(i)^2} $ for a vector $\vec{v}$ in $\bbR^{d}$.

\paragraph{Recap of Two Diversity Objectives}
\label{sec:pre:diversity}

We formally define two popular distance-based diversity objectives.
We assume that a pairwise distance $d(i,j)$ is given between every pair of items $i,j$.
One objective is the \emph{intra-list distance (ILD)},
which is defined as
\begin{align*}
 \ILD(S) \triangleq \frac{1}{{|S| \choose 2}} \sum_{i \neq j \in S} d(i,j)
\end{align*}
for an item set $S$.
The definition of ILD is intuitive, as it simply takes the average of the pairwise distances between all the items in $S$. 
The other is \emph{dispersion},
which is defined as the minimum pairwise distance between selected items:
\begin{align*}
 \disp(S) \triangleq \min_{i \neq j \in S} d(i,j).
\end{align*}
Dispersion is stricter than ILD in that 
it evaluates the pairwise distance among $S$ in the \emph{worst-case} sense.

We can flexibly choose from any distance function $d$ depending on the application.
Such a distance function is often a \emph{metric}; i.e.,
the following three axioms are satisfied for any items $i,j,k$:
(1)~identity of indiscernibles: $d(i,j) = 0 \iff i=j$;
(2)~symmetry: $d(i,j) = d(j,i)$;
(3)~triangle inequality: $d(i,j) + d(j,k) \geq d(i,k)$.
Commonly-used distance metrics in diversified recommendation include
the Euclidean distance \cite{ashkan2015optimal,sha2016framework},
i.e., $d(i,j) \triangleq \| \vec{x}_i - \vec{x}_j \|$, where
$\vec{x}_i$ and $\vec{x}_j$ are the feature vectors of items $i$ and $j$, respectively,
the cosine distance \cite{cheng2017learning,kaminskas2017diversity}, and
the Jaccard distance \cite{yu2009it,gollapudi2009axiomatic,kaminskas2017diversity}.

\begin{algorithm}[t] 
\caption{Greedy heuristic.}
\label{alg:greedy}
\small
\begin{algorithmic}[1]
    \Require
    diversity objective $\divf: 2^{[n]} \to \bbR_+$;
    \# items $k \in [n]$.
    \State \textbf{for } $\ell = 1 \To k$ \textbf{do }
    $i_\ell \leftarrow \argmax_{i \in [n] \setminus \{ i_1, \ldots, i_{\ell-1} \}} \divf(\{i_1, \ldots, i_{\ell-1}, i\}) $.
    \State \textbf{return} $\{ i_1, \ldots, i_k \}$.
\end{algorithmic}
\end{algorithm}

\paragraph{Greedy Heuristic}\label{sec:pre:greedy}
Here, we explain a greedy heuristic for enhancing diversity.
This heuristic has been frequently used in diversified recommendations,
and thus we use it for theoretical and empirical analyses of ILD and dispersion in \cref{sec:theory,sec:practice,sec:recommend}.

Consider the problem of selecting a set of $k$ items that maximize
the value of a particular diversity objective~$\divf$.
This problem is \textbf{NP}-hard, even if
$\divf$ is restricted to ILD \cite{tamir1991obnoxious} and dispersion \cite{ravi1994heuristic,erkut1990discrete}.
However, we can obtain an approximate solution to this problem using the simple greedy heuristic shown in \cref{alg:greedy}.
Given a diversity objective $\divf: 2^{[n]} \to \bbR_+$ on $n$ items and
an integer $k \in [n]$ representing the number of items to be recommended,
the greedy heuristic iteratively selects an item of $[n]$, not having been chosen so far,
that maximizes the value of $\divf$.
This heuristic has the following advantages from both theoretical and practical perspectives:
(1) it is \emph{efficient} because the number of evaluating $\divf$ is at most $nk$;
(2) it provably finds a $\frac{1}{2}$-approximate solution to
maximization of ILD \cite{birnbaum2009improved} and dispersion \cite{ravi1994heuristic},
which performs nearly optimal in practice.

\section{Theoretical Comparison}\label{sec:theory}

We present a theoretical analysis of the comparison between ILD and dispersion.
Our goal is to elucidate the \emph{correlation} between two diversity objectives.
Once we establish that enhancing a diversity objective $\divf$ results in
an increase in another $\divg$ to some extent,
we merely maximize $\divf$ to obtain
diverse items with respect to \emph{both} $\divf$ and $\divg$.
In contrast, if there is no such correlation,
we shall characterize what $\divf$ and $\divg$ are representing or enhancing.
The remainder of this section is organized as follows:
\cref{sec:theory:method} describes our analytical methodology,
\cref{sec:theory:results} summarizes our results, and
\cref{sec:theory:lessons} is devoted to lessons learned based on our results.

\subsection{Our Methodology}
\label{sec:theory:method}
We explain how to quantify the correlation between two diversity objectives.
Suppose we are given a diversity objective $\divf: 2^{[n]} \to \bbR_+$ over $n$ items and
an integer $k \in [n]$ denoting the output size (i.e., the number of items to be recommended).
We define \emph{$\divf$-diversification} as
the following optimization problem:
\begin{align*}
\max_{S \in {[n] \choose k}}\; \divf(S).
\end{align*}
Hereafter,
the optimal item set of $\divf$-diversification is denoted $S_{\divf,k}^*$
and the optimal value is denoted $\OPT_{\divf,k}$; namely, we define
$S_{\divf,k}^* \triangleq \argmax_{S \in {[n] \choose k}} \divf(S)$ and
$\OPT_{\divf,k} \triangleq \divf(S_{\divf,k}^*)$.
We also denote by $S_{\divf,k}^\Gr$ the set of $k$ items selected using the greedy heuristic on $\divf$.
We omit the subscript ``$k$'' when it is clear from the context.
Concepts related to approximation algorithms are also introduced.
\begin{definition}
We say that a $k$-item set $S$ is a \emph{$\rho$-approximation}
to $\divf$-diversification for some $\rho \leq 1$ if it holds that
\begin{align*}
    \divf(S) \geq \rho \cdot \OPT_{\divf,k}.
\end{align*}
Parameter $\rho$ is called the \emph{approximation factor}.
\end{definition}
For example, the greedy heuristic returns a $\frac{1}{2}$-approximation for 
\ILD-diversification; i.e.,
$\ILD(S_{\ILD}^\Gr) \geq \frac{1}{2} \cdot \OPT_{\ILD}$.

We now quantify the correlation 
between a pair of diversity objectives $\divf$ and $\divg$.
The primary logic is to think of
the optimal set $S_{\divf,k}^*$ for $\divf$-diversification
as an algorithm for $\divg$-diversification.
The correlation is measured using the approximation factor of this algorithm
for $\divg$-diversification, i.e.,
\begin{align}
\label{eq:theory:approx-factor}
\frac{\divg(S_{\divf,k}^*)}{\OPT_{\divg,k}}.
\end{align}
Intuitively, 
if this factor is sufficiently large,
then we merely maximize the value of $\divf$;
e.g., if \cref{eq:theory:approx-factor} is $0.99$,
then any item set having the optimum $\divf$ is also nearly-optimal with respect to $\divg$.
However, when \cref{eq:theory:approx-factor} is very low,
such an item set is not necessarily good with respect to $\divg$; namely,
$\divf$-diversification does \emph{not} imply $\divg$-diversification.
Note that we can replace $S_{\divf,k}^*$ with
the greedy solution, whose approximation factor is
$\frac{\divg(S_{\divf,k}^\Gr)}{\OPT_{\divg,k}}$.
Our analytical methodology is twofold:
\begin{itemize}[leftmargin=*]
\item[\textbf{1.}] We prove a guarantee on the approximation factor; i.e.,
there exists a factor $\rho$ such that
$\frac{\divg(S_{\divf}^*)}{\OPT_{\divg}} \geq \rho$ for \emph{every} set of items with a distance metric.
\item[\textbf{2.}] We
construct an input to indicate inapproximability; i.e.,
there exists a (small) factor $\rho'$ such that
$\frac{\divg(S_{\divf}^*)}{\OPT_{\divg}} < \rho'$
for \emph{some} item set with a distance metric.
Such an input demonstrates the case in which
$\divf$ and $\divg$ are quite different;
thus, we can use it to characterize what $\divf$ and $\divg$ represent.
\end{itemize}

\subsection{Our Results}
\label{sec:theory:results}
We now present our results,
each of which (i.e., a theorem or claim) is followed by a remark devoted to its intuitive implication.
Given that ILD and dispersion differ only in that
the former takes the \emph{average} and the latter the \emph{minimum} over all pairs of items,
an item set with a large dispersion value is expected to possess a large ILD value.
This intuition is first justified.
We define the \emph{diameter} $D$ for $n$ items
as the maximum pairwise distance; i.e.,
$
D \triangleq \max_{i \neq j \in [n]} d(i, j),
$
and denote by $d_k^*$ the maximum dispersion among $k$ items; i.e.,
$
d_k^* \triangleq \OPT_{\disp,k}.
$
Our first result is the following, whose proof is deferred to \cref{app:proof}.
\begin{theorem}\label{thm:ILD-disp-approx}
The following inequalities hold for any input and distance metric:
$
    \frac{\ILD(S_{\disp,k}^*)}{\OPT_{\ILD,k}} \geq \frac{d_k^*}{D} \text{   and   }
    \frac{\ILD(S_{\disp,k}^\Gr)}{\OPT_{\ILD,k}} \geq \max\left\{\frac{d_k^*}{2D}, \frac{1}{k} \right\}.
$
In other words,
the optimal size-$k$ set to \disp-diversification is a
$\frac{d_k^*}{D}$-approximation to \ILD-diversification, and
\cref{alg:greedy} on \disp returns
a $\max\{\frac{d_k^*}{2D}, \frac{1}{k}\}$-approximation to \ILD-diversification.
\end{theorem}

\noindent
\emph{\textbf{Remark:}}
\cref{thm:ILD-disp-approx} implies that
\emph{the larger the dispersion, the larger the ILD},
given that $D$ is not significantly large.
In contrast,
if the maximum dispersion $d_k^*$ is much smaller than $D$,
the approximation factor $\frac{d_k^*}{D}$ becomes less fascinating.
Fortunately, the greedy heuristic exhibits a $\frac{1}{k}$-approximation,
which facilitates a data-independent guarantee.

We demonstrate that \cref{thm:ILD-disp-approx} is almost tight,
whose proof is deferred to \cref{app:proof}.

\begin{claim}
\label{clm:ILD-disp-tight}
There exists an input such that
the pairwise distance is the Euclidean distance between feature vectors, and
the following holds:
$
    \frac{\ILD(S_{\disp,k}^*)}{\OPT_{\ILD,k}} = \bigO\left(\frac{d_k^*}{D}\right)
    \text{ and }
    \frac{\ILD(S_{\disp,k}^\Gr)}{\OPT_{\ILD,k}} = \bigO\left(\frac{1}{k} + \frac{d_k^*}{D}\right).
$
In particular, \cref{thm:ILD-disp-approx} is tight up to constant.
\end{claim}

\noindent
\emph{\textbf{Remark:}}
The input used in the proof of \cref{clm:ILD-disp-tight}
consists of two ``clusters'' such that
the intra-cluster distance of each cluster is extremely small (specifically, $\epsilon$) and
the inter-cluster distance between them is large.
The ILD value is maximized when the same number of items from each cluster are selected.
However,
any set of three or more items has a dispersion $\epsilon$; namely,
we cannot distinguish between
the largest-ILD case and the small-ILD case based on the value of dispersion.

In the reverse direction, 
we provide a very simple input such that
\emph{no matter how large the ILD value is, the dispersion value can be $0$},
whose proof is deferred to \cref{app:proof}.
\begin{claim}
\label{clm:disp-ILD-inapprox}
There exists an input such that the pairwise distance is
the Euclidean distance
and
$
    \frac{\disp(S_{\ILD}^*)}{\OPT_{\disp}} =
    \frac{\disp(S_{\ILD}^\Gr)}{\OPT_{\disp}} = 0.
$
In other words,
greedy or exact maximization of \ILD does not have any approximation guarantee to \disp-diversification.
\end{claim}
\noindent
\emph{\textbf{Remark:}}
The input used in the proof of \cref{clm:disp-ILD-inapprox}
consists of (duplicates allowed) points on a line segment.
Dispersion selects distinct points naturally.
In contrast,
ILD prefers points on the two ends of the segment,
which are redundant.

\subsection{Lessons Learned}
\label{sec:theory:lessons}

\begin{figure}[t]
    \centering 
    \null\hfill
    \subfloat[Two separated circles (cf.~\cref{clm:ILD-disp-tight}).\label{fig:theory:lesson1}]{\includegraphics[width=.5\hsize]{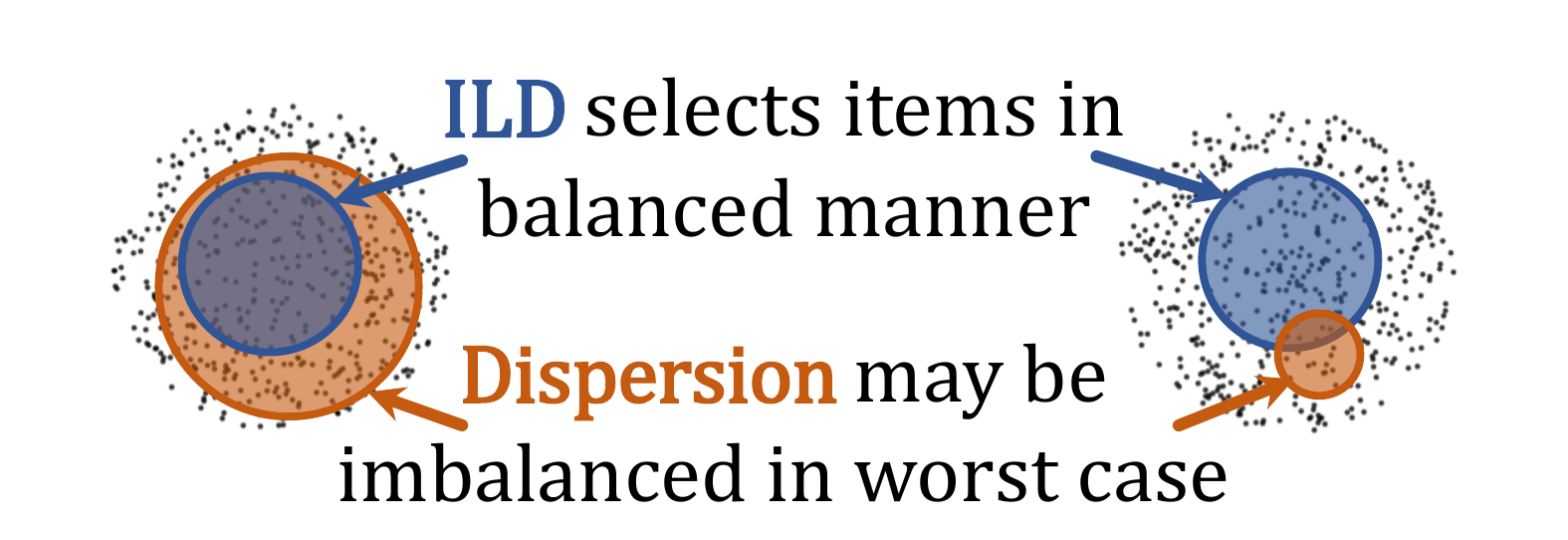}}%
    \hfill
    \subfloat[An ellipse (cf.~\cref{clm:disp-ILD-inapprox}).\label{fig:theory:lesson2}]{\includegraphics[width=.5\hsize]{fig/lesson2.pdf}}%
    \hfill\null
    \caption{Two inputs for which maximization of ILD and dispersion results in very different solutions.}
    \label{fig:theory:lessons}
\end{figure}

Based on the theoretical investigations so far,
we discuss the pros and cons of ILD and dispersion.
\cref{fig:theory:lessons} shows two illustrative inputs such that
maximization of ILD and dispersion results in very different solutions,
where each item is a $2$-dimensional vector and
the distance between items is measured by the Euclidean distance.

\begin{itemize}[leftmargin=*]
\item \textbf{Pros of ILD}:
If the entire item set is separated into two ``clusters'' as shown in \cref{fig:theory:lesson1},
ILD selects items in a well-balanced manner; i.e.,
a nearly equal number of items from each cluster are chosen (supported by \cref{clm:ILD-disp-tight}).

\item \textbf{Cons of ILD}:
ILD may select duplicate items that are very close (or even identical) to each other.
Suppose that we are given feature vectors in an ellipse shown in \cref{fig:theory:lesson2}.
Then, ILD would select items from the left and right ends,
each of which consists of similar feature vectors (supported by \cref{clm:disp-ILD-inapprox});
even more, items in the middle of the ellipse are never chosen.

In practice,
if item features are given by \emph{dense} vectors
such as those generated by deep neural networks,
ILD is undesirable because it selects many nearly-identical vectors.

\item \textbf{Pros of dispersion}:
If the entire item set is ``well-dispersed'' as in \cref{fig:theory:lesson2},
then so are the items chosen by dispersion as well.

\item \textbf{Cons of dispersion}:
Dispersion may overlook distant item pairs that would have contributed to ILD.
Suppose that we are given feature vectors in two circles in \cref{fig:theory:lesson1}.
Because the dispersion value of any (three or more) items is small
whereas the diameter is large,
we cannot distinguish distant items from close items
using only the dispersion value.
Thus, dispersion may select items in an unbalanced manner in the worst case
(as in \cref{clm:ILD-disp-tight}).

In practice,
if item features are given by \emph{sparse} (e.g., 0-1) vectors, such as 
indicator functions defined by genre or topic information,
dispersion may not be favorable, because its value becomes $0$ whenever 
two or more items with the same feature are selected.
\end{itemize}

\section{Gaussian Intra-List Distance}
\label{sec:gauss}

In \cref{sec:theory:lessons},
we discussed that ILD and dispersion have their own \emph{extreme behaviors}.
We now argue that they can be viewed as limits in the sense of a kernel function over items, i.e.,
we apply the Gaussian kernel to ILD.
The \emph{Gaussian kernel}
for two vectors $\vec{x},\vec{y} \in \bbR^d$ is defined as
$
K(\vec{x}, \vec{y}) \triangleq \exp\Bigl(-\frac{\|\vec{x} - \vec{y}\|^2}{2 \sigma^2}\Bigr),
$
where $\sigma > 0$ is a \emph{bandwidth} parameter
that controls the smoothness of the estimated function in kernel methods.
Since the kernel function can be considered as \emph{similarity score},
we can define the \emph{kernel distance} \cite{phillips2011gentle} as
$ d_K(\vec{x},\vec{y}) = \sqrt{2 -2 K(\vec{x},\vec{y})}. $
Using this kernel distance, we define the \emph{Gaussian ILD (GILD)} as
\begin{align}
    \GILD_\sigma(S)
    \triangleq
    \frac{1}{{|S| \choose 2}} \sum_{i \neq j \in S} \sqrt{2- 2 \exp\left(-\frac{d(i,j)^2}{2 \sigma^2}\right)},
\end{align}
where $d$ is a distance metric and $\sigma$ is a bandwidth parameter.\footnote{
Note that we have replaced the Euclidean distance in
$\exp\Bigl(-\frac{\|\vec{x}_1 - \vec{x}_j\|^2}{2\sigma^2}\Bigr)$ by $d$ so that we can use any distance metric.
}
The following asymptotic analysis shows that GILD interpolates ILD and dispersion,
whose proof is deferred to \cref{app:proof}.

\begin{theorem}\label{thm:GILD}
GILD approaches ILD as the value of $\sigma$ goes to $\infty$, and
it approaches dispersion as the value of $\sigma$ goes to $0$ (up to scaling and addition by a constant).
\end{theorem}

\cref{thm:GILD} implies that GILD behaves as a compromise
between ILD and dispersion by tuning the bandwidth parameter $\sigma$:
the value of $\sigma$ must be small if
we do not want the selected items to be close to each other;
$\sigma$ must be large if we want to include (a few) distance items.

We use GILD to better understand the empirical behavior of ILD and dispersion.
In particular, we are interested to know whether
GILD can avoid the extreme behavior of ILD and dispersion.

\subsection{Choosing the Value of $\sigma$}
\label{sec:gauss:sigma}

Here, we briefly establish how to choose the value of $\sigma$
in \cref{sec:practice}.
As will be shown in \cref{sec:practice:results:sigma},
GILD usually exhibits extreme behaviors like ILD or dispersion.
We wish to determine the value of $\sigma$ for which GILD interpolates them.
Suppose that we have selected $k$ items, denoted $S$.
In \cref{eq:GILD:expand} in the proof of \cref{thm:GILD},
for the first two terms to be dominant, 
we must have
$
C \gg ({k \choose 2}-C) \cdot \epsilon_\sigma,
$
which implies that
$\sigma \gg \sqrt{\frac{(\disp(S)+\delta)^2 - \disp(S)^2}{2 \log ({k \choose 2} - 1)}}.$
Based by this,
we propose the following two schemes for determining the value of $\sigma$,
referred to as the \emph{adjusted minimum} and the \emph{adjusted median}:
\begin{align}
\label{eq:GILD:sigma-val}
    \sigma_S^{\min} \triangleq
    \frac{\min_{i \neq j \in S} d(i,j)}{\sqrt{2 \log ({k \choose 2} - 1)}} \text{ and }
    \sigma_S^{\med} \triangleq
    \frac{\median_{i \neq j \in S} d(i,j)}{\sqrt{2 \log ({k \choose 2} - 1)}}.
\end{align}
Note that $\sigma_S^{\min} \leq \sigma_S^{\med}$, and the adjusted median
mimics the median heuristic \cite{gretton2012optimal,garreau2017large} in kernel methods.
In \cref{sec:practice}, we empirically justify that
dividing by $\sqrt{2 \log ({k \choose 2} - 1)}$ is necessary.
Since $\sigma_S^{\min}$ and $\sigma_S^{\med}$ depend on $S$,
we run the greedy heuristic while adjusting the value of $\sigma$ \emph{adaptively}
using \cref{eq:GILD:sigma-val}:
More precisely,
in line~1 of \cref{alg:greedy}, we define
$f(\{i_1, \ldots, i_\ell, i\}) \triangleq \GILD_{\sigma}(S \cup \{i\}) - \GILD_{\sigma}(S) $, where
$S \triangleq \{i_1, \ldots, i_\ell\}$ and
$\sigma$ is $\sigma_{S \cup \{i\}}^{\min}$ or $\sigma_{S \cup \{i\}}^{\med}$.
We further slightly modify this heuristic so that
it selects the pair of farthest items when $k=2$
because $\sqrt{2 \log ({k \choose 2} - 1)}$ is $-\infty$.

\begin{table*}[t]
    \centering
    \small
    \begin{minipage}[t]{.32\textwidth}
    \renewcommand{\arraystretch}{0.8}
    \center
    \begin{tabular}{c|lll} \toprule
    \multirow{2}{*}{\diagbox{$\divf$}{$\divg$}} & \multicolumn{3}{c}{\textbf{rel.~score} $\divg(S_{\divf,k}^\Gr) / \divg(S_{\divg,k}^\Gr)$} \\
    & \ILD & \disp & \GILD \\ \midrule
    \ILD & -- & 0.424 & 0.997 \\
    \disp & 0.941 & -- & 1.000 \\
    \GILD & \textbf{0.972} & \textbf{0.818} & -- \\
    \Random & 0.345 & 0.053 & 0.934 \\
    \bottomrule
    \end{tabular}
    \caption{Average rel.~score of each pair of diversity objs.~for \feedback on \ML.}
    \label{tab:rel_ml_svd}
    \end{minipage}
    \hfill
    \begin{minipage}[t]{.32\textwidth}
    \renewcommand{\arraystretch}{0.8}
    \center
    \begin{tabular}{c|lll} \toprule
    \multirow{2}{*}{\diagbox{$\divf$}{$\divg$}} & \multicolumn{3}{c}{\textbf{rel.~score} $\divg(S_{\divf,k}^\Gr) / \divg(S_{\divg,k}^\Gr)$} \\
    & \ILD & \disp &\GILD \\ \midrule
    \ILD & -- & 0.211 & 0.999 \\
    \disp & 0.975 & -- & 0.998 \\
    \GILD & \textbf{0.997} & \textbf{0.360} & -- \\
    \Random & 0.142 & 0.001 & 0.810 \\
    \bottomrule
    \end{tabular}
    \caption{Average rel.~score of each pair of diversity objs.~for \feedback on \Ama.}
    \label{tab:rel_ama_svd}
    \end{minipage}
    \hfill
    \begin{minipage}[t]{.32\textwidth}
    \renewcommand{\arraystretch}{0.8}
    \center
    \begin{tabular}{c|lll} \toprule
    \multirow{2}{*}{\diagbox{$\divf$}{$\divg$}} & \multicolumn{3}{c}{\textbf{rel.~score} $\divg(S_{\divf,k}^\Gr) / \divg(S_{\divg,k}^\Gr)$} \\
    & \ILD & \disp & \GILD \\ \midrule
    \ILD & -- & 0.048 & 0.797 \\
    \disp & 0.859 & -- & 0.936 \\
    \GILD & \textbf{0.889} & \textbf{0.195} & -- \\
    \Random & 0.842 & 0.162 & 0.955 \\
    \bottomrule
    \end{tabular}
    \caption{Average rel.~score of each pair of diversity objs.~for \circles.}
    \label{tab:rel_circles}
    \end{minipage}
    \\
    \begin{minipage}[t]{.32\textwidth}
    \renewcommand{\arraystretch}{0.8}
    \center
    \begin{tabular}{c|lll} \toprule
    \multirow{2}{*}{\diagbox{$\divf$}{$\divg$}} & \multicolumn{3}{c}{\textbf{rel.~score} $\divg(S_{\divf,k}^\Gr) / \divg(S_{\divg,k}^\Gr)$} \\
    & \ILD  & \disp & \GILD \\ \midrule
    \ILD & -- & 0.153 & 0.976 \\
    \disp & 0.959 & -- & 0.999 \\
    \GILD & \textbf{0.970} & \textbf{0.911} & -- \\
    \Random &  0.877 & 0.000 & 0.933 \\
    \bottomrule
    \end{tabular}
    \caption{Average rel.~score of each pair of diversity objs.~for \genre on \ML.}
    \label{tab:rel_ml_genre}
    \end{minipage}
    \hfill
    \begin{minipage}[t]{.32\textwidth}
    \renewcommand{\arraystretch}{0.8}
    \center
    \begin{tabular}{c|lll} \toprule
    \multirow{2}{*}{\diagbox{$\divf$}{$\divg$}} & \multicolumn{3}{c}{\textbf{score} $\divg(S_{\divf,k}^\Gr) / \divg(S_{\divg,k}^\Gr)$} \\
    & \ILD & \disp& \GILD \\ \midrule
    \ILD & -- & 0.378 & 0.996 \\
    \disp & 0.979 & -- & 0.999 \\
    \GILD & \textbf{0.989} & \textbf{0.926} & -- \\ 
    \Random & 0.966 & 0.137 & 0.990 \\
    \bottomrule
    \end{tabular}
    \caption{Average rel.~score of each pair of diversity objs.~for \genre  on \Ama.}
    \label{tab:rel_ama_genre}
    \end{minipage}
    \hfill
    \begin{minipage}[t]{.32\textwidth}
    \renewcommand{\arraystretch}{0.8}
    \center
    \begin{tabular}{c|lll} \toprule
    \multirow{2}{*}{\diagbox{$\divf$}{$\divg$}} & \multicolumn{3}{c}{\textbf{rel.~score} $\divg(S_{\divf,k}^\Gr) / \divg(S_{\divg,k}^\Gr)$} \\
    & \ILD & \disp& \GILD \\ \midrule
    \ILD & -- & 0.041 & 0.652 \\
    \disp & 0.684 & -- & 1.000 \\
    \GILD & \textbf{0.758} & \textbf{0.272} & -- \\ 
    \Random & 0.567 & 0.185 & 0.985 \\
    \bottomrule
    \end{tabular}
    \caption{Average rel.~score of each pair of diversity objs.~for \ellipse.}
    \label{tab:rel_ellipse}
    \end{minipage}
\end{table*}

\begin{figure*}[t]
\null\hfill
\begin{minipage}{0.32\linewidth}%
    \centering
    \includegraphics[width=0.9\hsize]{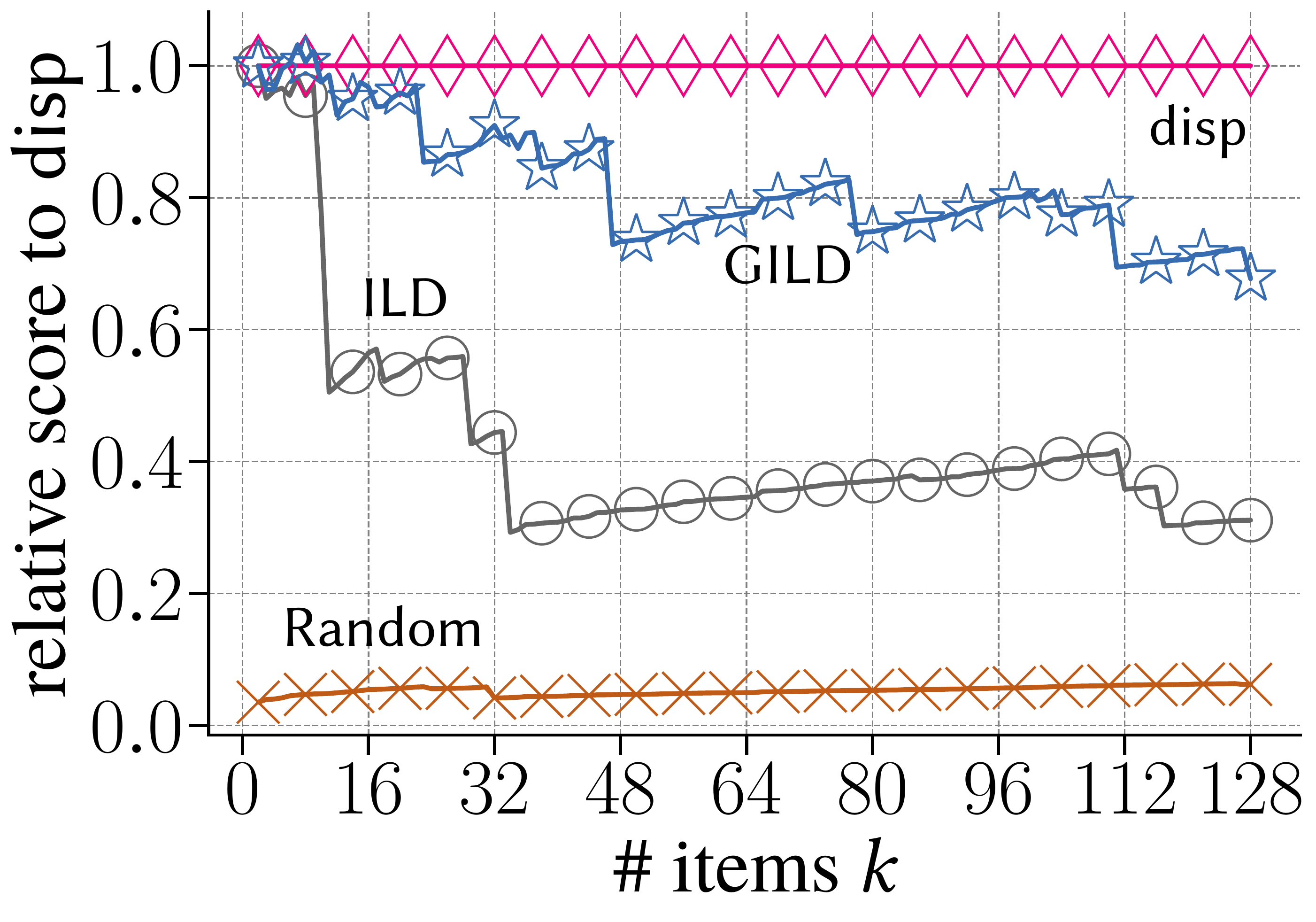}
    \caption{Relative score of each objective to dispersion for \feedback on \ML.}
    \label{fig:rel_ml_svd_disp}
\end{minipage}%
\hfill
\begin{minipage}{0.32\linewidth}%
    \centering
    \includegraphics[width=0.9\hsize]{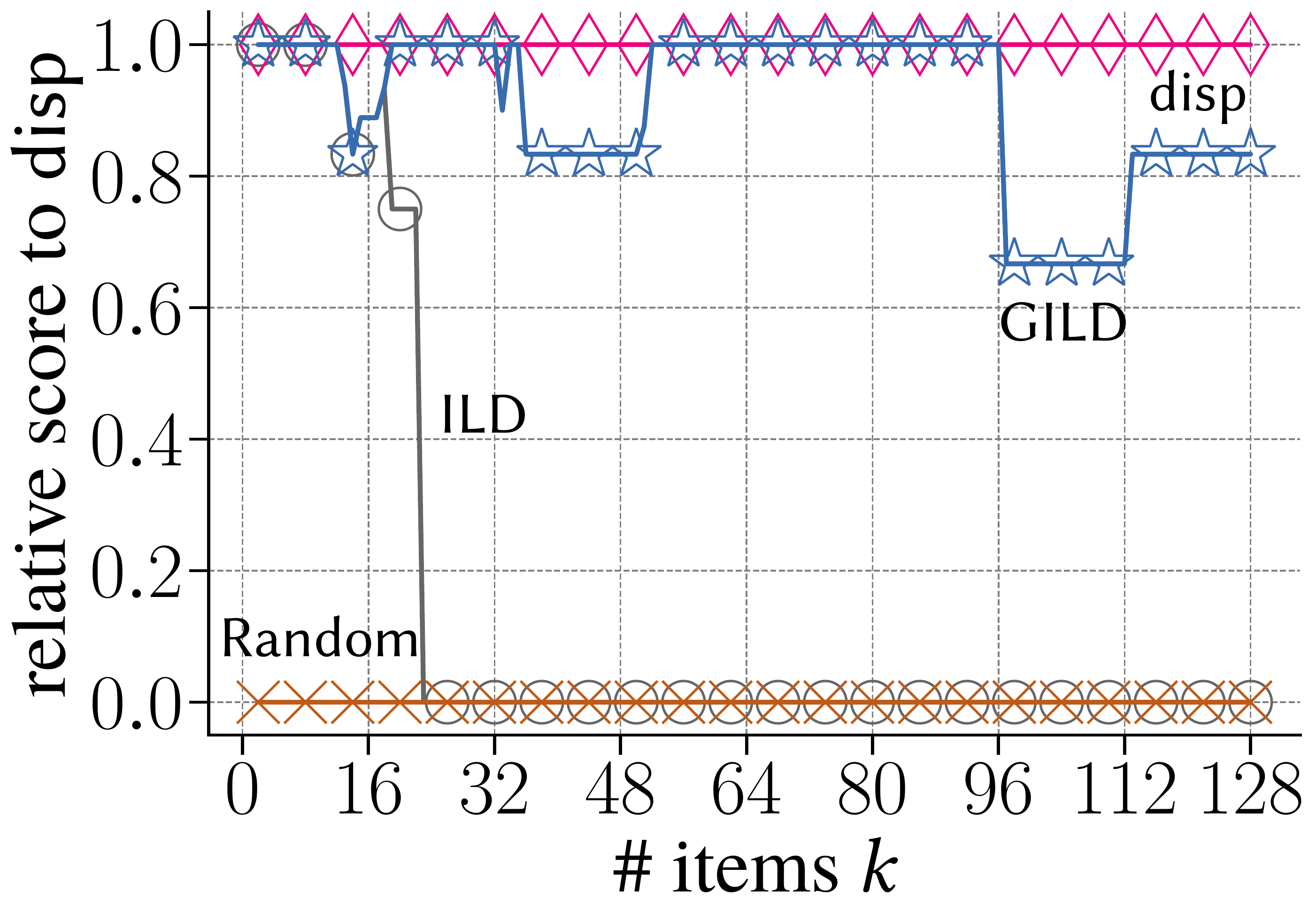}
    \caption{Relative score of each objective to dispersion for \genre on \ML.}
    \label{fig:rel_ml_genre_disp}
\end{minipage}%
\hfill
\begin{minipage}{0.32\linewidth}%
    \centering
    \includegraphics[width=0.9\hsize]{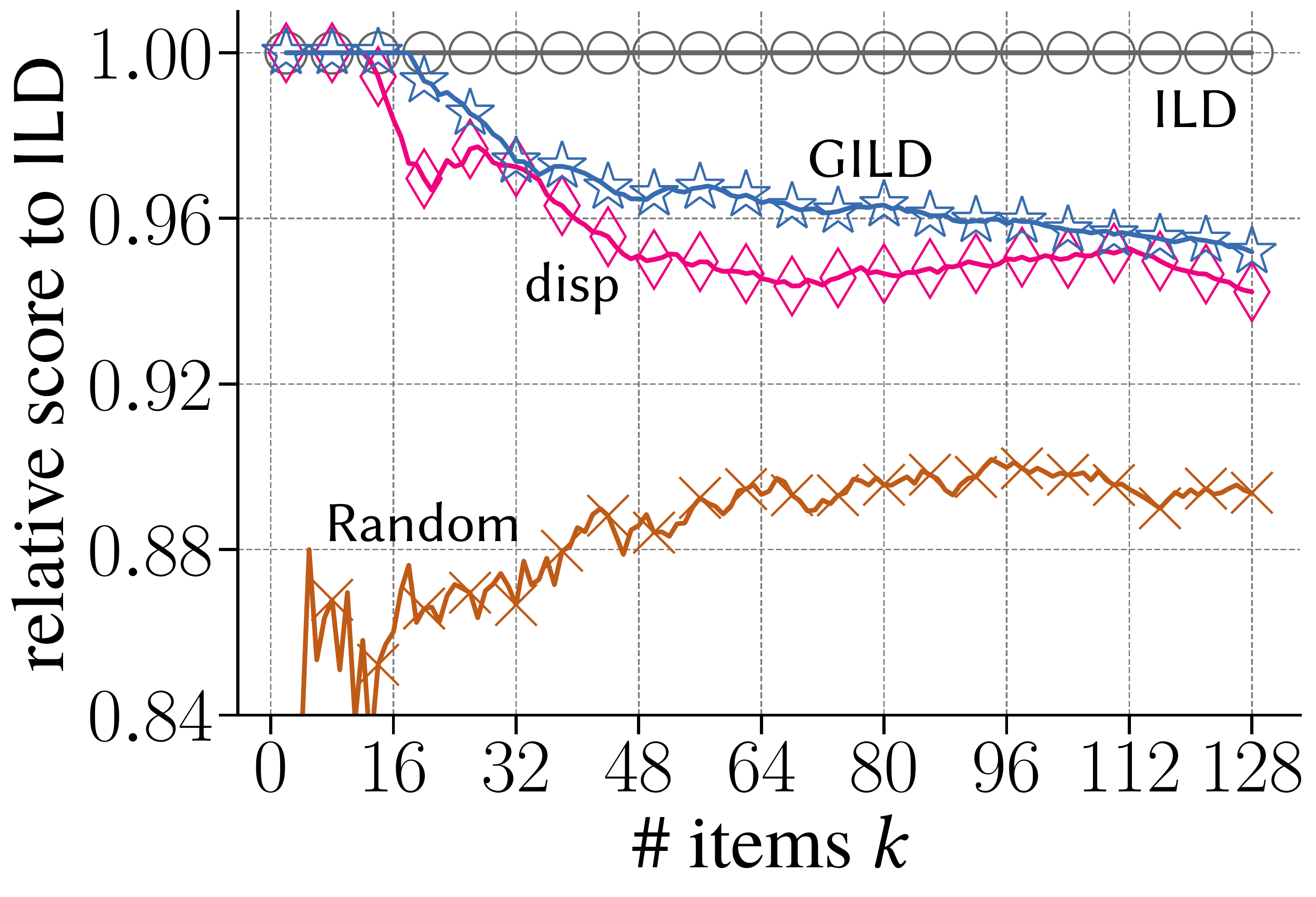}
    \caption{Relative score of each objective to ILD for \genre on \ML.}
    \label{fig:rel_ml_genre_ILD}
\end{minipage}%
\hfill\null
\end{figure*}

\section{Empirical Comparison}
\label{sec:practice}

We report the experimental results of the empirical comparison among the diversity objectives analyzed in \cref{sec:theory,sec:gauss}.
The theoretical results in \cref{sec:theory} demonstrate that
each objective captures its own notion of diversity; thus,
enhancing one objective is generally unhelpful in improving another.
One may think that such results based on \emph{worst-case analysis} are too pessimistic to be applied in practice; for instance,
ILD may be used to enhance dispersion in real data,
even though any positive approximation guarantee is impossible.
Thus, we \emph{empirically} analyze the approximation factor for the diversity objectives examined thus far.

\subsection{Settings}
\label{sec:practice:settings}
\subsubsection{Datasets}
We use two real-world datasets including feedback and genre information and two synthetic datasets.
\begin{itemize}[leftmargin=*]
    \item[\textbf{1.}] \textbf{MovieLens 1M} (\ML)
    \cite{harper2015movielens,movielensurl}:
    Genre information is associated with each movie; there are $ 18$ genres.
    We extracted the subset in which users and movies have at least $20$ ratings,
    resulting in $995$ thousand ratings on $3{,}000$ movies from $6{,}000$ users.
    \item[\textbf{2.}] \textbf{Amazon Review Data Magazine Subscriptions} (\Ama)
    \cite{ni2019justifying,amazonurl}:
    Each product contains categorical information, and there are $165$ categories.
    We extracted the subset in which 
    all users and movies have at least five ratings,
    resulting in $4{,}200$ reviews of $720$ products from $664$ users.
    \item[\textbf{3.}] \textbf{Random points in two separated circles} (\circles, \cref{fig:theory:lesson1}):
    Consist of $1{,}000$ random points in two circles whose
    radius is $\frac{1}{4}$ and centers are $-\frac{3}{4}$ and $\frac{3}{4}$.
    \item[\textbf{4.}] \textbf{Random points in an ellipse} (\ellipse, \cref{fig:theory:lesson2}):
    Consist of $1{,}000$ random points in an ellipse of flattening $\frac{3}{4}$.
\end{itemize}

\subsubsection{Distance Metrics}
We use two types of distance metrics for real-world datasets.
\begin{itemize}[leftmargin=*]
\item[\textbf{1.}]
\emph{Implicit feedback} (\feedback for short):
Let $\mat{X}$ be a user-item implicit feedback matrix over $m$ users and $n$ items,
such that $X_{u,i}$ is $1$ if user $u$ interacts with item $i$, and
$0$ if there is no interaction.
We run singular value decomposition on $\mat{X}$ with dimension $d \triangleq 32$
to obtain
$\mat{X} = \mat{U}\mat{\Sigma}\mat{V}^\top$, where
$\mat{V}^\top = [\vec{v}_1, \ldots, \vec{v}_n] \in \bbR^{d \times n} $.
The feature vector of item $i$ is then defined as $\vec{v}_i$ and 
the distance between two items $i,j$ is given by the Euclidean distance $d(i,j) \triangleq \| \vec{v}_i - \vec{v}_j \|$.
\item[\textbf{2.}]
\emph{Genre information} (\genre for short):
We denote by $G_i$ the set of genres that item $i$ belongs to.
The distance between two items $i,j$ is given by the Jaccard distance
$d(i,j) \triangleq 1 - \frac{|G_i \cap G_j|}{|G_i \cup G_j|}$.
Multiple items may have the same genre set;
i.e., $d(i,j) = 0$ for some $i \neq j$.
\end{itemize}
For two synthetic datasets, we simply use the Euclidean distance.

\subsubsection{Diversity Enhancement Algorithms}
We apply the greedy heuristic (\cref{alg:greedy}) to
ILD, dispersion, and GILD with the adjusted median.
A baseline that returns a random set of items (denoted \Random) is implemented.
Experiments were conducted on a Linux server with an Intel Xeon 2.20GHz CPU and 62GB RAM.
All programs were implemented using Python~3.9.

\subsection{Results}
\label{sec:practice:results}
We calculate the empirical approximation factor for each pair of diversity objectives $\divf$ and $\divg$ as follows.
First, we run the greedy heuristic on $\divf$ to extract up to $128$ items.
The empirical approximation factor of $\divf$ to $\divg$ is obtained by
$
\divg(S_{\divf,k}^\Gr) / \divg(S_{\divg,k}^\Gr)
$
for each $k \in [128]$.
This factor usually takes a number from $0$ to $1$ and is simply referred to as the \emph{relative score of $\divf$ to $\divg$}.
Unlike the original definition in \cref{eq:theory:approx-factor},
we do not use $\OPT_{\divg,k}$ because
its computation is \textbf{NP}-hard.
\cref{tab:rel_ml_svd,tab:rel_ml_genre,tab:rel_ama_svd,tab:rel_ama_genre,tab:rel_circles,tab:rel_ellipse}
report the average relative score over $k = 2, \ldots, 128$.

\subsubsection{ILD vs.~Dispersion vs.~GILD in Practice}

The relative score of ILD to dispersion is first investigated, where
we proved that no approximation guarantee is possible (\cref{clm:disp-ILD-inapprox}).
In almost all cases, the relative score is extremely low, with the highest being $0.424$.
This is because that
multiple items with almost-the-same features were selected,
resulting in a small (or even $0$) value of dispersion.
\cref{fig:rel_ml_svd_disp} shows that
ILD selects items that have similar feature vectors when $k = 34$;
we thus confirmed the claim in \cref{sec:theory:lessons} that
ILD selects nearly-identical items in the case of dense feature vectors.
Moreover,
\cref{fig:rel_ml_genre_disp} shows that
it selects duplicate items that share the same genre set
at $k = 23$.

We then examine the relative score of dispersion to ILD,
for which we provided an approximation factor of $\max\{\frac{d_k^*}{2D}, \frac{1}{k}\}$ (\cref{thm:ILD-disp-approx}).
\cref{tab:rel_ml_svd,tab:rel_ml_genre,tab:rel_ama_svd,tab:rel_ama_genre,tab:rel_circles,tab:rel_ellipse} show that
the relative score is better than $0.859$ except for \ellipse,
which is better than expected from $\frac{1}{k}$.
\cref{fig:rel_ml_genre_ILD} also indicates that the relative score does \emph{not} decay significantly;
e.g., at $k=100$, the relative score is better than $0.94$ even though
the worst-case approximation factor is $\frac{1}{k} = 0.01$.

It is evident that GILD has a higher relative score to ILD than dispersion, and
a higher relative score to dispersion than ILD for all settings.
That is, GILD finds an intermediate set between ILD and dispersion,
suggesting that ILD and dispersion exhibit the extreme behavior in practice as discussed in \cref{sec:theory}.

\begin{figure*}[t]
    \centering
    \null\hfill
    \subfloat[$128$ points chosen by ILD.\label{fig:plot_ellipse_ILD}]{\includegraphics[width=.25\textwidth]{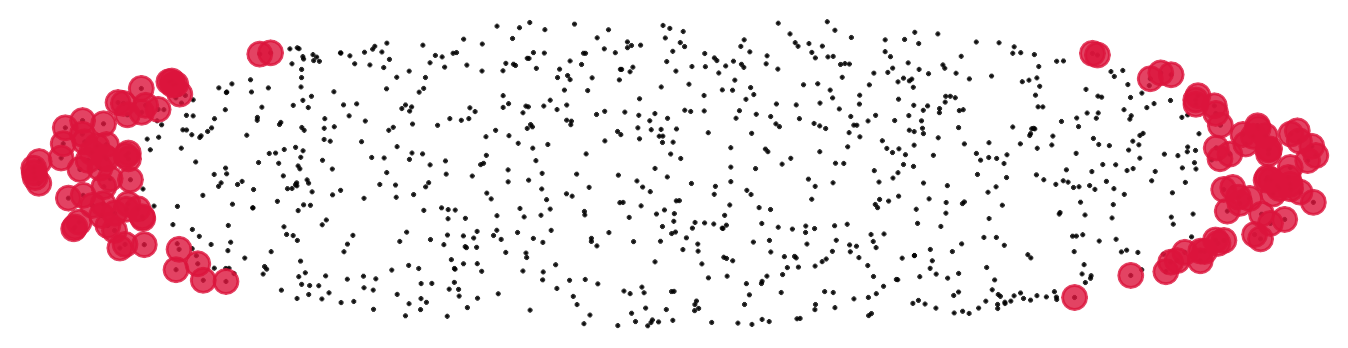}}%
    \hfill
    \subfloat[$128$ points chosen by dispersion.\label{fig:plot_ellipse_disp}]{\includegraphics[width=.25\textwidth]{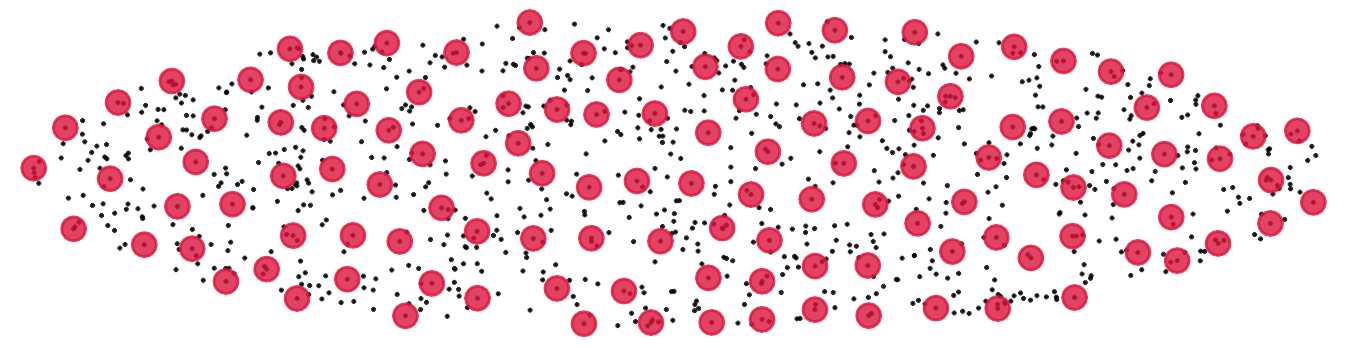}}%
    \hfill
    \subfloat[$128$ points chosen by GILD.\label{fig:plot_ellipse_GILD}]{\includegraphics[width=.25\textwidth]{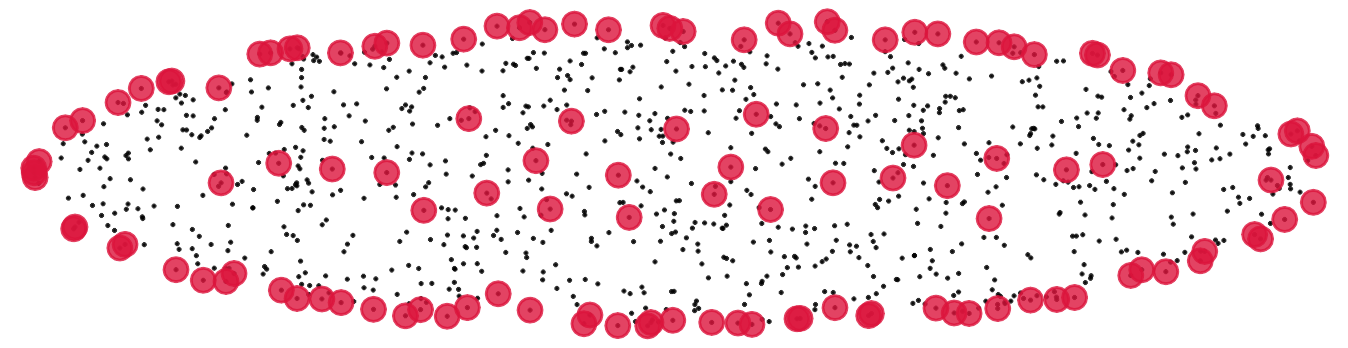}}%
    \hfill\null
    \caption{
    $128$ points (big red circles) of \ellipse selected by greedily maximizing each objective with the Euclidean distance.}
    \label{fig:plot_ellipse}
\end{figure*}%

\begin{figure*}[t]
    \centering
    \null\hfill
    \subfloat[$128$ points chosen by ILD.\label{fig:plot_circles_ILD}]{\includegraphics[width=.25\textwidth]{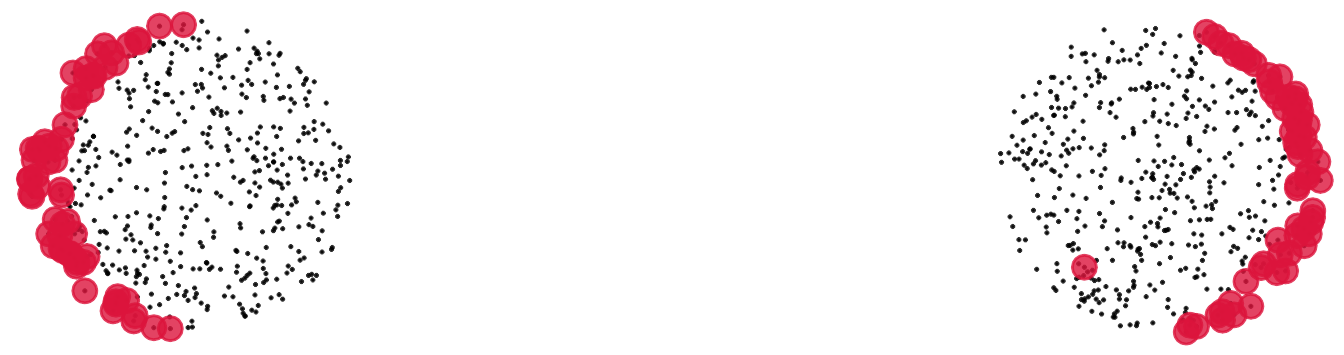}}%
    \hfill
    \subfloat[$128$ points chosen by dispersion.\label{fig:plot_circles_disp}]{\includegraphics[width=.25\textwidth]{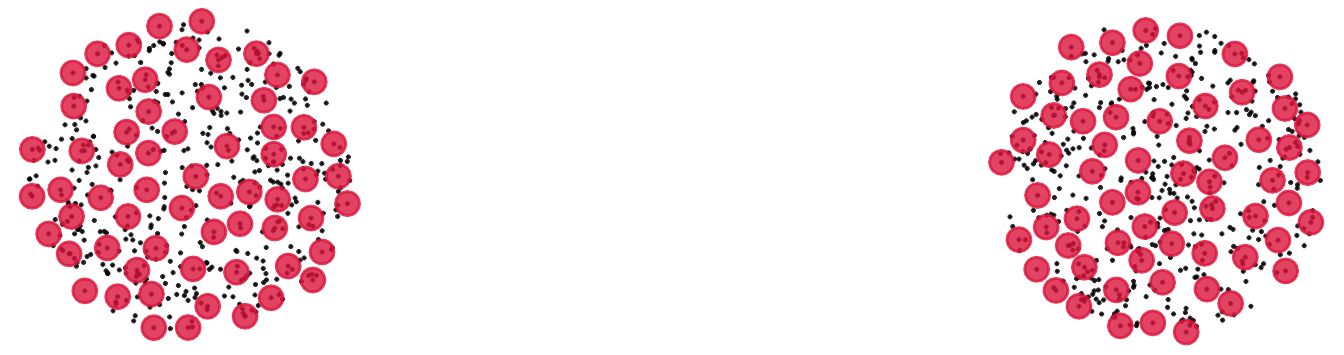}}%
    \hfill
    \subfloat[$128$ points chosen by GILD.\label{fig:plot_circles_GILD}]{\includegraphics[width=.25\textwidth]{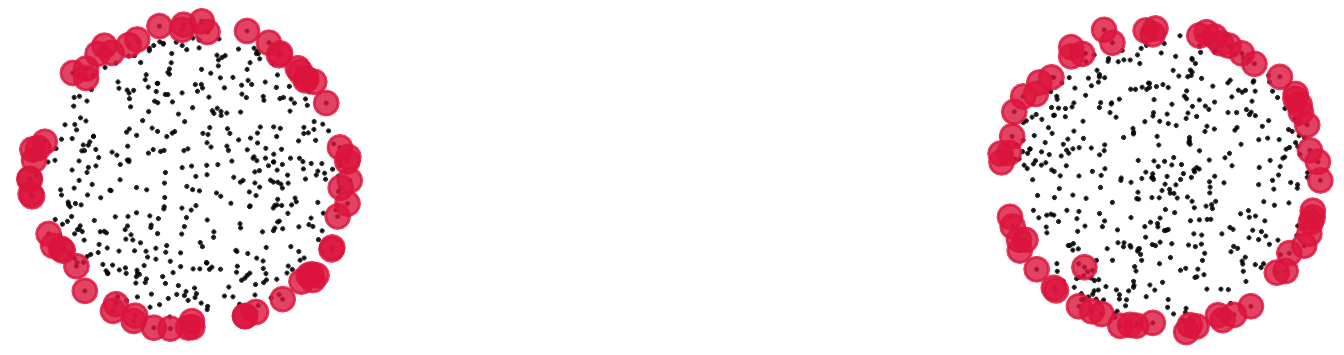}}%
    \hfill\null
    \caption{
    $128$ points (big red circles) of \circles selected by greedily maximizing each objective with the Euclidean distance.}
    \label{fig:plot_circles}
\end{figure*}

\begin{wrapfigure}[11]{r}{0.23\textwidth}
\centering
\includegraphics[width=\hsize]{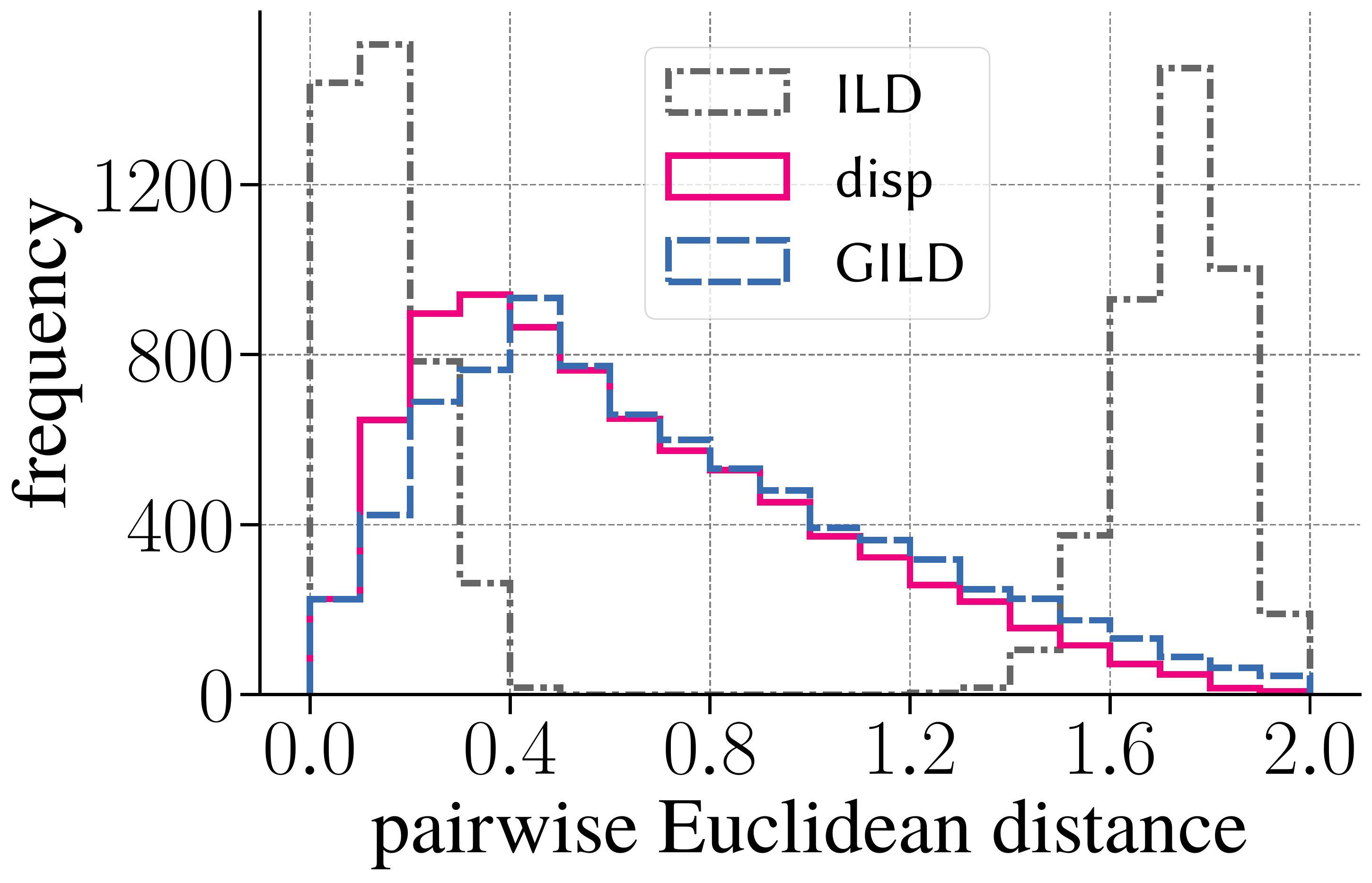}
\caption{Histogram of the pairwise distances of the selected items on \ellipse.}
\label{fig:hist_ellipse}
\end{wrapfigure}
\subsubsection{Qualitative Analysis via Visualization}
\label{sec:practice:results:vis}
We qualitatively assess the diversity objectives based on the visualization of synthetic datasets.
We first investigate \ellipse,
in which ILD may select duplicate items (see \cref{sec:theory:lessons}).
\cref{fig:plot_ellipse} shows
items of \ellipse that are selected by each diversity objective;
\cref{fig:hist_ellipse} shows the histogram of
the pairwise Euclidean distances between the selected items.
The items selected by ILD can be partitioned into two groups: the left and right ends of the ellipse (\cref{fig:plot_ellipse_ILD}).
The histogram further shows that
the inter-group distance between them is approximately $1.8$
whereas the intra-group distance is close to $0$.
Thus, the drawback of ILD in \cref{sec:theory:lessons} occurs empirically.
Unlike ILD,
the items selected by dispersion are well dispersed (\cref{fig:plot_ellipse_disp});
however, it misses many pairs of distant items as shown in \cref{fig:hist_ellipse}.
One reason for this result is
given that dispersion is the minimum pairwise distance,
maximizing the value of dispersion does not lead to the selection of distant item pairs,
as discussed in \cref{sec:theory:lessons}.
In contrast, the items chosen by GILD are not only scattered (\cref{fig:plot_ellipse_GILD});
they include more dissimilar items than dispersion, as shown in the histogram.
This observation can be explained by the GILD mechanism,
which takes the sum of the kernel distance over \emph{all} pairs.

We then examine \circles.
\cref{fig:plot_circles} shows that 
each diversity objective selects almost the same number of items from each cluster.
In particular, the potential drawback of dispersion discussed in \cref{sec:theory:lessons},
i.e., the imbalance of selected items in the worst case, does not occur empirically.

\begin{figure}[t]
    \centering%
    \null\hfill
    \subfloat[$k=16$.]{\includegraphics[width=.49\hsize]{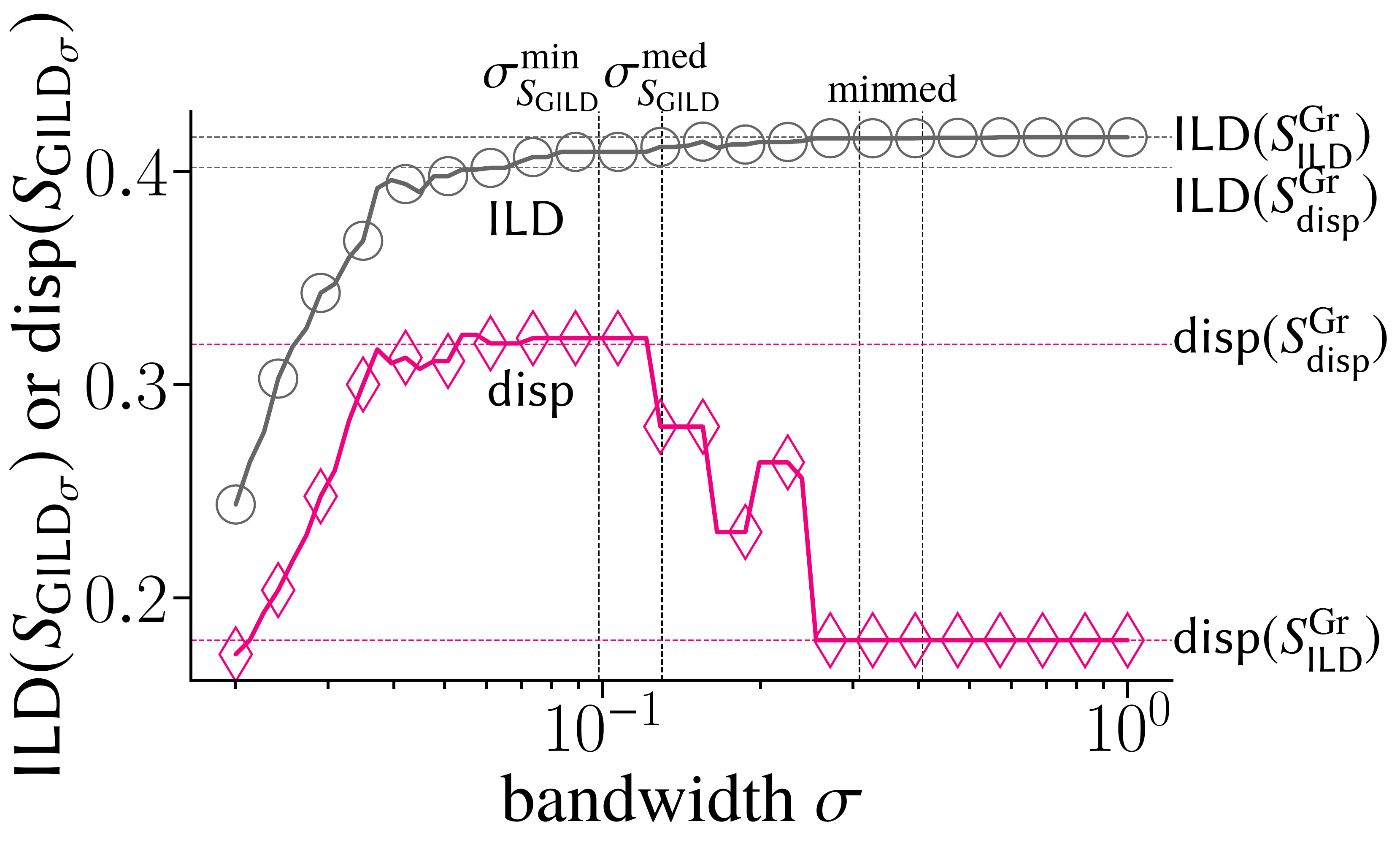}}%
    \hfill
    \subfloat[$k=128$.]{\includegraphics[width=.49\hsize]{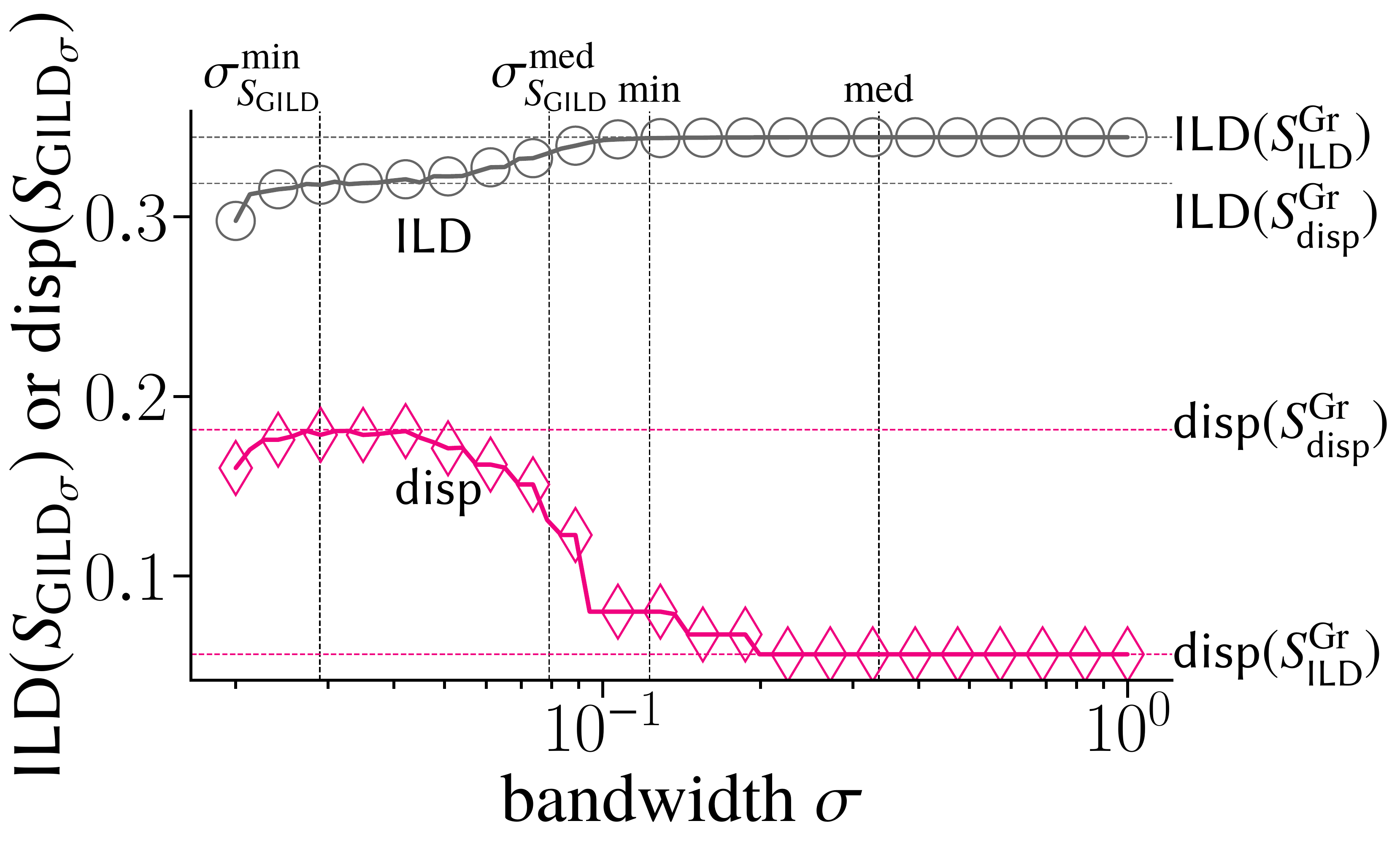}}%
    \hfill\null
    \caption{Trade-off between ILD and dispersion.
    For each value of $\sigma$, we plot $\ILD(S_{\GILD_\sigma,k}^\Gr)$ and $\disp(S_{\GILD_\sigma,k}^\Gr)$.
    }
    \label{fig:trade_ml_svd}
\end{figure}

\subsubsection{Investigation of the Effect of $\sigma$ on GILD}
\label{sec:practice:results:sigma}
We investigate the empirical effect of the value of $\sigma$ on the behavior of GILD.
Specifically, we examine how GILD interpolates between ILD and dispersion by changing $\sigma$,
as suggested in \cref{thm:GILD}.
Setting the value of $\sigma$ to
each of $64$ equally-spaced numbers on a log scale from $0.02$ to $1$,
we greedily maximize $\GILD_\sigma$ for \feedback on \ML
to obtain a $k$-item set $S_{\GILD_\sigma,k}$.
We also run the adaptive greedy heuristic,
which is oblivious to the value of $\sigma$,
to obtain a $k$-item set $S_{\GILD,k}$.
\cref{fig:trade_ml_svd} plots values of ILD and dispersion for
each obtained set $S_{\GILD_\sigma,k}$
of size $k=16,128$.
The vertical lines correspond to
the adjusted minimum $\sigma_{S_{\GILD,k}}^{\min}$,
adjusted median $\sigma_{S_{\GILD,k}}^{\med}$,
minimum $\min_{i \neq j \in S_{\GILD,k}} d(i,j)$, and 
median $\median_{i \neq j \in S_{\GILD,k}} d(i,j)$.
Horizontal lines correspond to
$\ILD(S_{\ILD}^\Gr) \approx \OPT_{\ILD}$, $\ILD(S_{\disp}^\Gr)$,
$\disp(S_{\disp}^\Gr) \approx \OPT_{\disp}$, and $\disp(S_{\ILD}^\Gr)$.
Observe first that
\ILD is monotonically increasing in $\sigma$ and approaches
$\OPT_{\ILD}$;
\disp is approximately decreasing in $\sigma$ and attains $\OPT_{\disp}$
for a ``moderately small'' value of $\sigma$,
which coincides with \cref{thm:GILD}.

Observe also that
the degradation of both \ILD and \disp occurs for small values of $\sigma$.
The reason
is that each term $\exp\Bigl(-\frac{d(i,j)^2}{2 \sigma^2}\Bigr)$ in GILD becomes
extremely small, causing a floating-point rounding error.
Setting $\sigma$ to the minimum and median results in a dispersion value of $\disp(S_{\ILD}^\Gr)$ when $k = 16$; i.e., the obtained set is almost identical to $S_{\ILD}^\Gr$.
In contrast,
setting $\sigma = \sigma_{S_{\GILD,k}}^{\min}$ is similar to $S_{\disp}^\Gr$;
setting $\sigma = \sigma_{S_{\GILD,k}}^{\med}$ yields
a set
whose dispersion is between $\disp(S_{\disp,k}^\Gr)$ and $\disp(S_{\ILD,k})$ and
whose ILD is in the middle of $\ILD(S_{\ILD,k}^\Gr)$ and $\ILD(S_{\disp,k})$.
Thus, using the adjusted median, and division by $\sqrt{2 \log{k \choose 2}-1}$ is crucial
for avoiding trivial sets.

\begin{wrapfigure}[10]{r}{0.23\textwidth}
\centering
\includegraphics[width=\hsize]{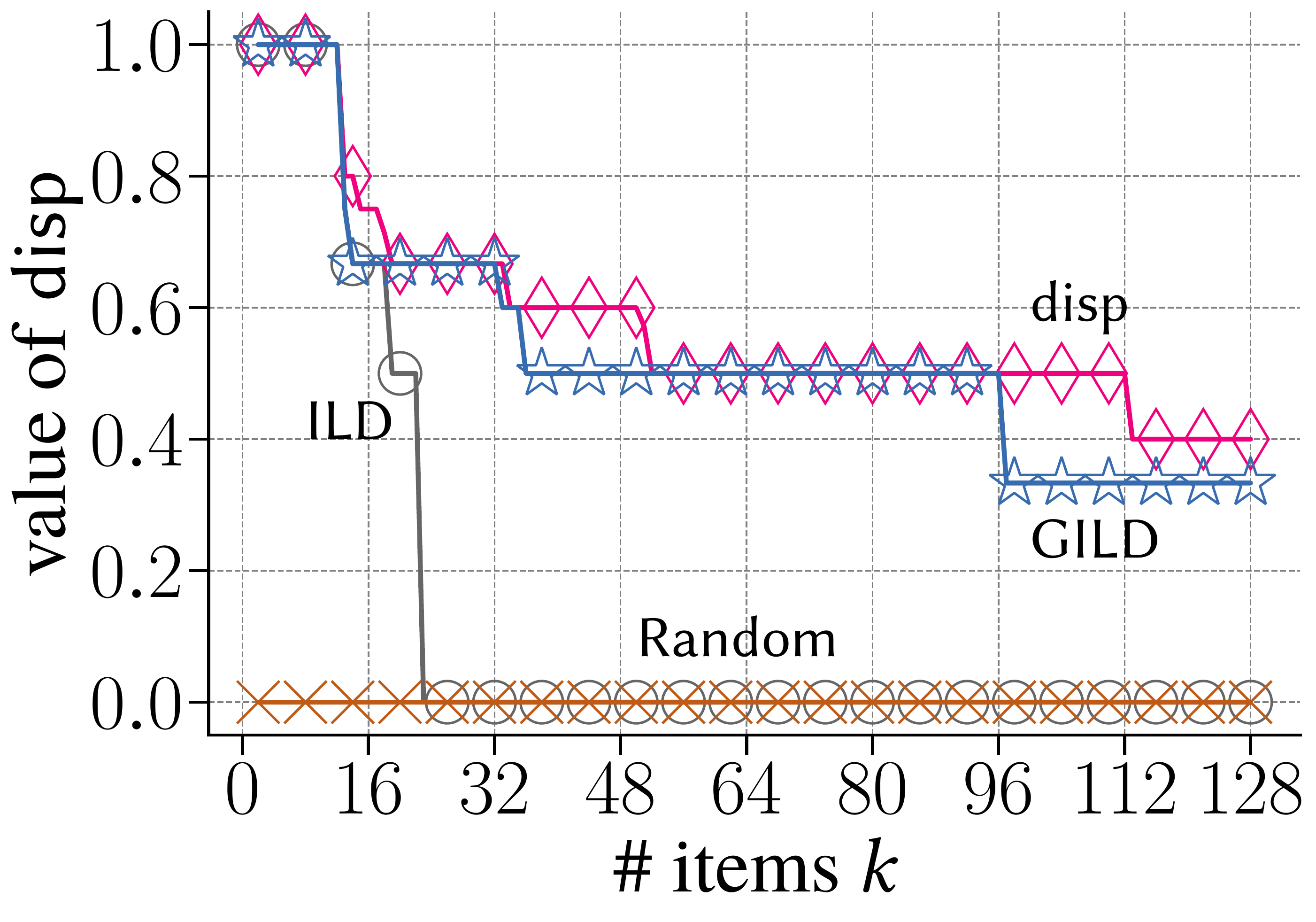}
\caption{
    Dispersion of items for \genre on \ML.
}
\label{fig:abs_ml_genre_disp}
\end{wrapfigure}
\begin{figure*}[ht]
    \centering
    \null\hfill
    \includegraphics[width=.3\textwidth]{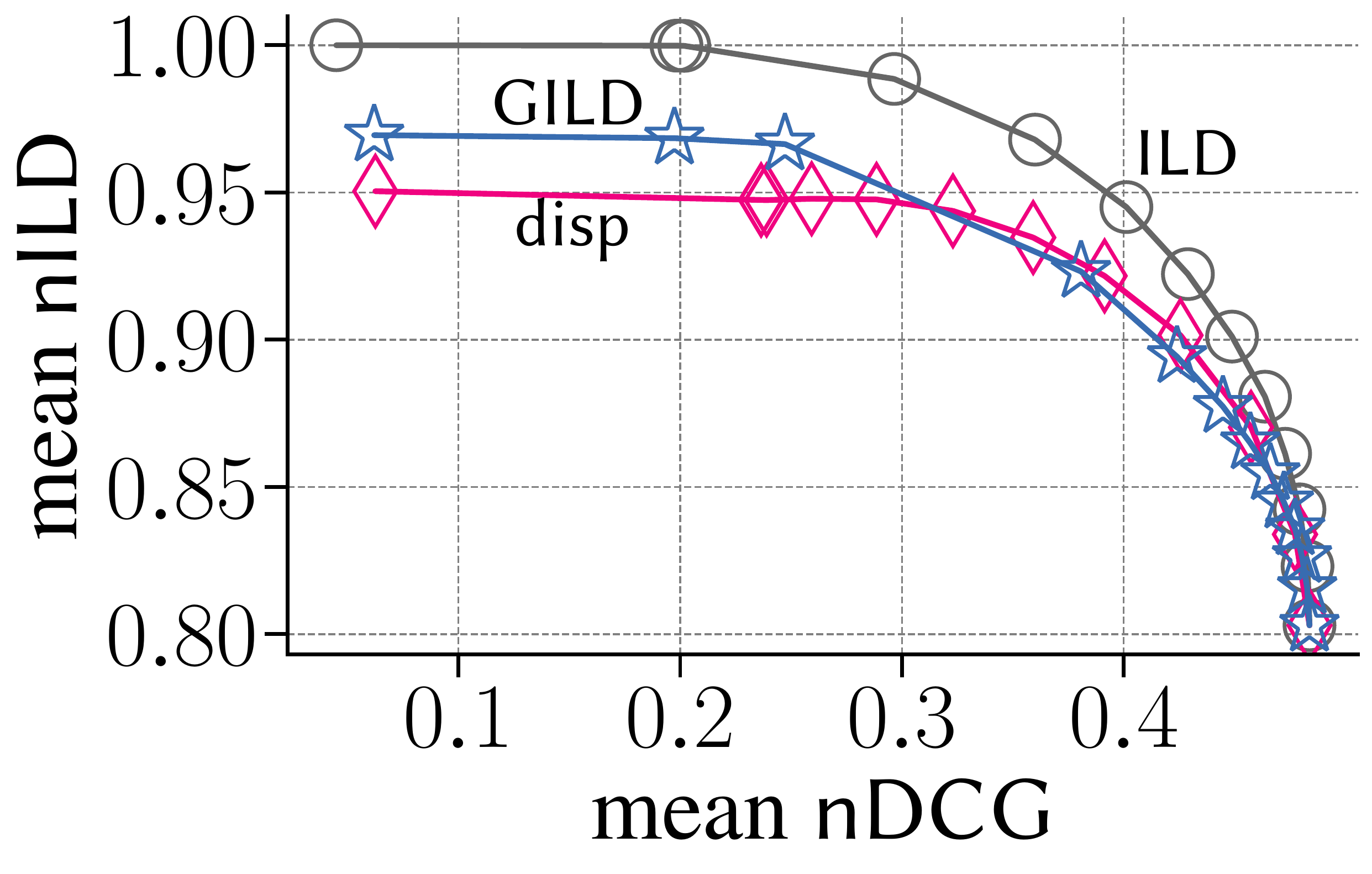}%
    \hfill%
    \includegraphics[width=.3\textwidth]{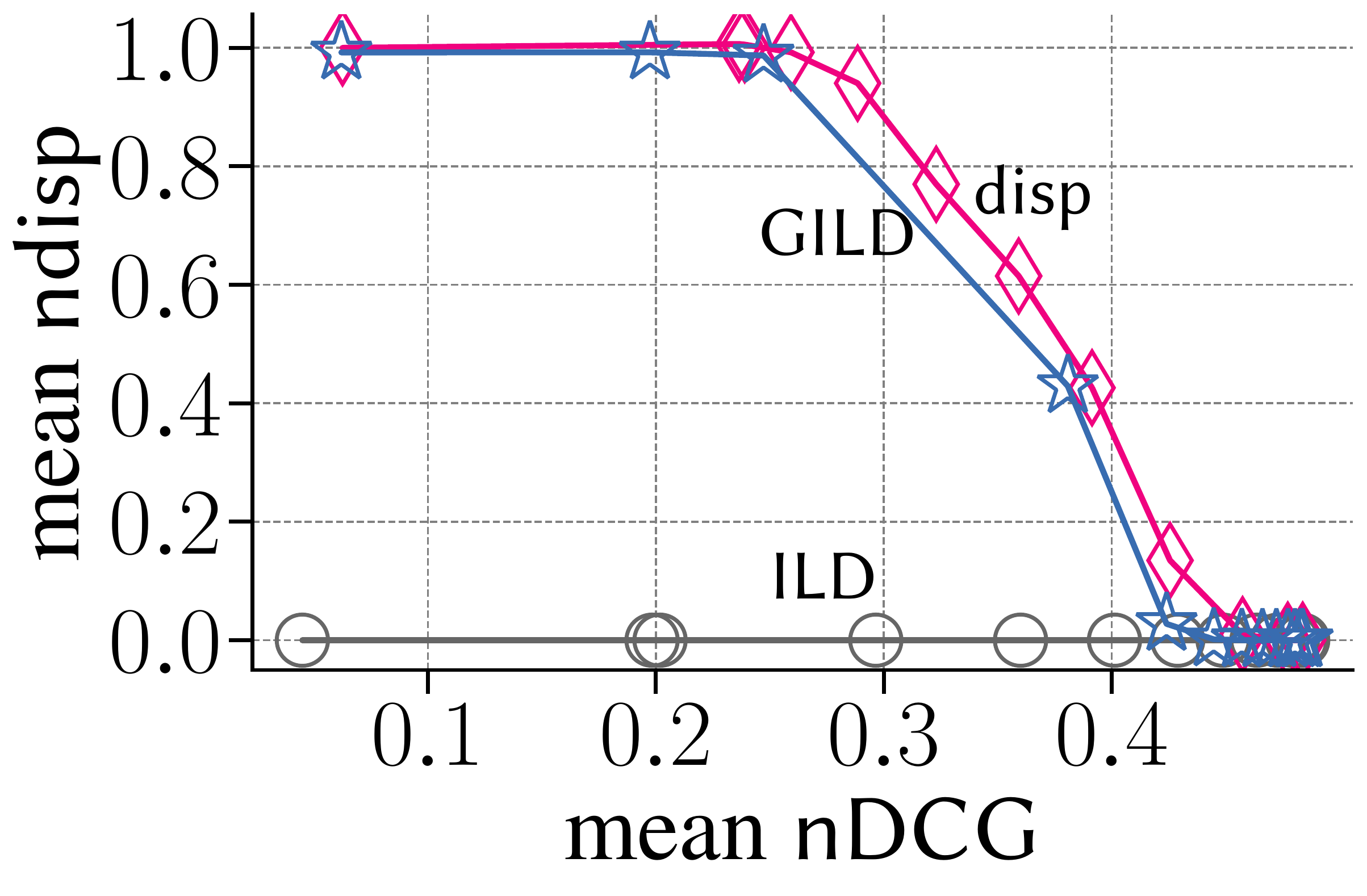}%
    \hfill%
    \includegraphics[width=.3\textwidth]{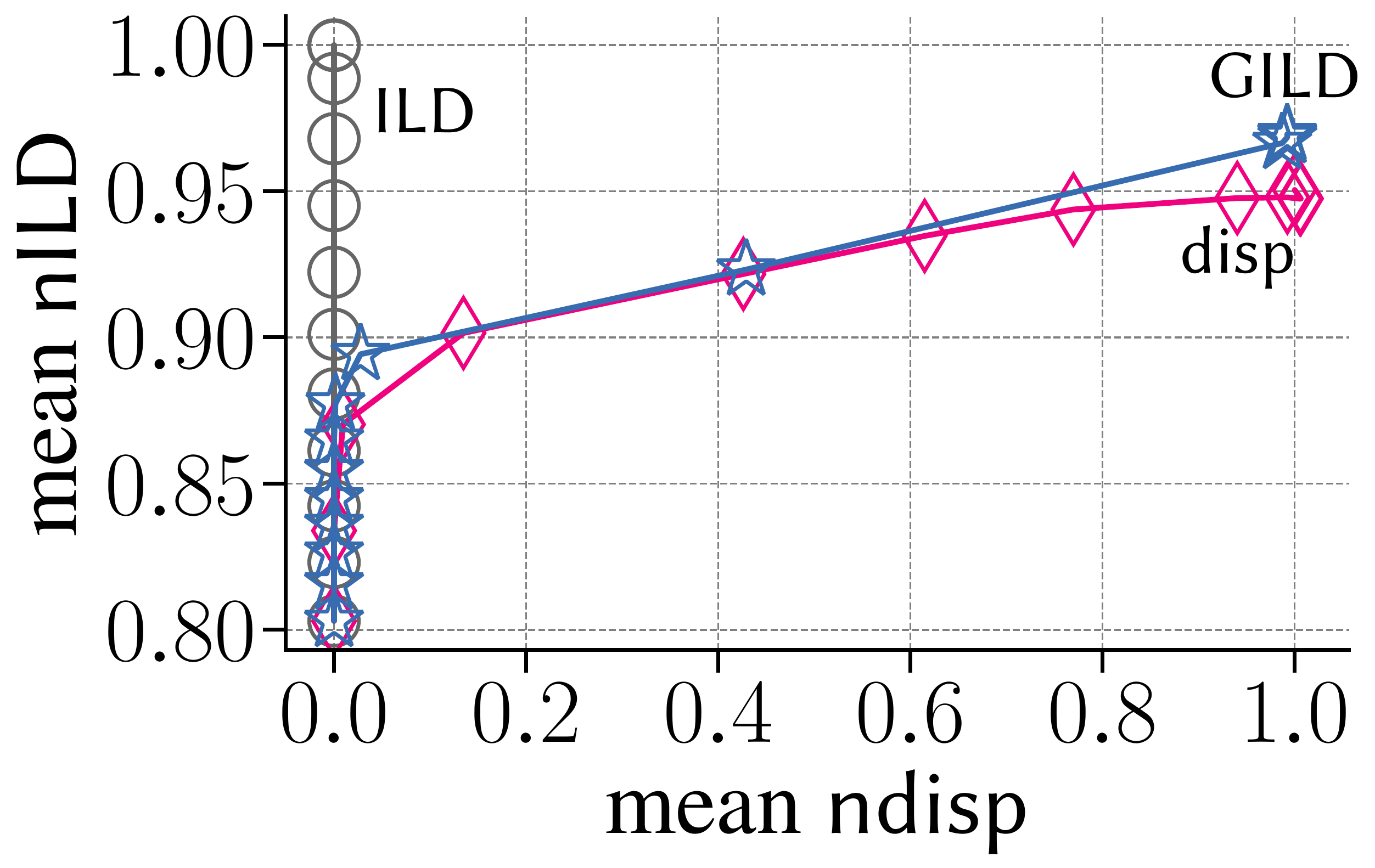}%
    \hfill\null
    \caption{
        Relation between each pair of \nDCG, \nILD, and \ndisp with regard to a trade-off parameter $\lambda$.
    }
    \label{fig:recommend}
\end{figure*}

\subsection{Discussions}
\label{sec:practice:discuss}
We discuss the empirical behavior of ILD, dispersion, and GILD.
Arguably, ILD easily selects many items that are similar or identical.
As shown in \cref{fig:plot_ellipse_ILD},
the chosen items are biased toward two distant groups, and
items in the middle of the two groups never appear.
This is undesirable if we wish to recommend very few items.

Such drawbacks of ILD can be resolved via dispersion.
Greedy maximization of dispersion also empirically enhances the ILD value.
However,
it may overlook distant item pairs, as discussed in \cref{sec:practice:results:vis}.
We also note that dispersion is not suitable for measuring  diversity.
As shown in \cref{fig:abs_ml_genre_disp},
the value of dispersion drops to nearly $0$ when selecting a moderate number of items;
it does \emph{not} return to a positive value.
Due to this nature,
dispersion may not be used to compare large item sets.

The empirical result of GILD implies that
ILD and dispersion are not appropriate for improving and/or evaluating distance-based diversity.
GILD partially circumvents the issues caused by the extreme behavior of ILD and dispersion,
thereby achieving the \emph{sweet spot} between them.
On the one hand, GILD extracts dissimilar items such that the dispersion value does not drop to $0$.
On the other hand,
GILD can select more dissimilar items than dispersion.
Similar to dispersion,
GILD cannot be used to \emph{compare} the diversity among distinct sets,
as shown in \cref{tab:rel_circles}, which indicates that
even \Random can have the highest GILD value.
This is because GILD with the adjusted median is designed to evaluate
the next item to be selected given a \emph{fixed} set of already-selected items.
To sum up, GILD works successfully
as an optimization objective interpolating ILD and dispersion and
as a tool for analyzing them empirically.

\section{Diversified Recommendation Results}
\label{sec:recommend}
Having a better understanding of the behavior of diversity objectives from both theoretical (\cref{sec:theory}) and empirical perspectives (\cref{sec:practice}),
we incorporate them into the recommendation methods.

\subsection{Settings}
\subsubsection{Dataset}
To investigate results 
produced by a recommendation method
using ILD, dispersion, and GILD,
we use the \ML dataset, the details of which are described in \cref{sec:practice:settings}.
We extracted the subset in which
users and movies have at least $20$ and $100$ ratings, respectively,
resulting in $370$ thousand ratings on $2{,}000$ movies from
$2{,}800$ users.
The obtained subset was further split into
training, validation, and test sets in a $60/20/20$ ratio 
according to weak generalization; i.e.,
they may not be disjoint in terms of users.

\subsubsection{Algorithms}
\label{sec:recommend:settings:algorithms}
We adopt Embarrassingly Shallow AutoEncoder (\easer) \cite{steck2019embarrassingly}
to estimate the predictive score $\mathsf{rel}_u(i)$
for item $i$ by user $u$ from a user-item implicit feedback matrix.
\easer has a hyperparameter for $L_2$-norm regularization, and
its value is tuned using the validation set.
We construct a distance metric based on the \emph{implicit feedback} in \cref{sec:practice:settings} to 
define ILD, dispersion, and GILD.
We then apply the greedy heuristic to a linear combination of relevance and diversity.
Specifically, given a set $S_{\ell-1}$ of already selected $\ell-1$ items,
we select the next item $i_{\ell}$ that maximizes the following objective:
\begin{align}
\label{eq:recommend:greedy}
    \mathsf{F}_{u,\divf,\lambda}(i) \triangleq 
    (1-\lambda) \cdot \mathsf{rel}_u(i) + \lambda \cdot \{ \divf(S_{\ell-1} \cup \{i\}) - \divf(S_{\ell-1}) \},
\end{align}
where $\lambda \in (0,1)$ is a trade-off parameter between relevance and diversity.
We run the greedy heuristic for
each $\divf$,
each value of $\lambda = 0, 0.1, 0.2, \ldots, 0.9, 0.99, 0.999, 1$, and
each user $u$
to retrieve a list of $k \triangleq 50$ items to be recommended to $u$,
denoted $S_{u,\divf,\lambda}$.
Experiments were conducted on the same environment as described in \cref{sec:practice}.

\subsubsection{Evaluation}
We evaluate the accuracy and diversity of the obtained sets as follows.
Let $R_u$ denote the set of relevant items to user $u$
(i.e., those interacting with $u$) in the test set.
We calculate the \emph{normalized Discounted Cumulative Gain (nDCG)} by
\begin{align*}
    \nDCG@k(S_{u,\divf,\lambda}; R_u) & \triangleq
    \Bigl( \sum_{\ell \in [\min\{k, |R_u|\}]} \frac{1}{\log_2(\ell+1)} \Bigr)^{-1} \cdot \\
    & \sum_{\ell \in [k]} \frac{[\!\![\ell\text{-th ranked item of }S_{u,\divf,\lambda} \text{ is in } R_u]\!\!]}{\log_2(\ell+1)}.
\end{align*}
We calculate the normalized versions of ILD and dispersion as
$\nILD(S_{u,\divf,\lambda}) \triangleq \frac{\ILD(S_{u,\divf,\lambda})}{\ILD(S_{u,\ILD}^\Gr)}$
and
$\ndisp(S_{u,\divf,\lambda}) \triangleq \frac{\disp(S_{u,\divf,\lambda})}{\disp(S_{u,\disp}^\Gr)}$,
respectively,
where $S_{u,\divf}^\Gr$ is the set of $k$ items obtained by
greedily maximizing $\divf$ on the set of items that do not appear
in the training or validation set.
We then take the mean
of \nDCG, \nILD, and \ndisp over all users.

\subsection{Results}
\cref{fig:recommend} shows the relation between each pair of \nDCG, \nILD, and \ndisp.
First, we observe a clear trade-off relationship between relevance and diversity regarding $\lambda$.
In particular, when diversity is not introduced into the objective (i.e., $\lambda = 0$), the mean \ndisp takes $0$,
which implies
that for most users, two or more of selected items have the same genre set.
As shown in \cref{sec:practice},
incorporating ILD does not avoid the case of $\ndisp = 0$.
In contrast,
dispersion and GILD with a moderate value of $\lambda$ enhance \nILD and \ndisp without substantially sacrificing accuracy.
Comparing dispersion and GILD,
it is observed that GILD achieves a slightly higher \nILD than dispersion:
When the mean \nDCG is close to $0.25$,
the means of \nILD for GILD and dispersion are $0.966$ and $0.948$, respectively, and
the means of \ndisp for them are $0.987$ and $0.992$, respectively.

\begin{figure}[tbp]
    \centering
    \null\hfill
    \includegraphics[width=.48\hsize]{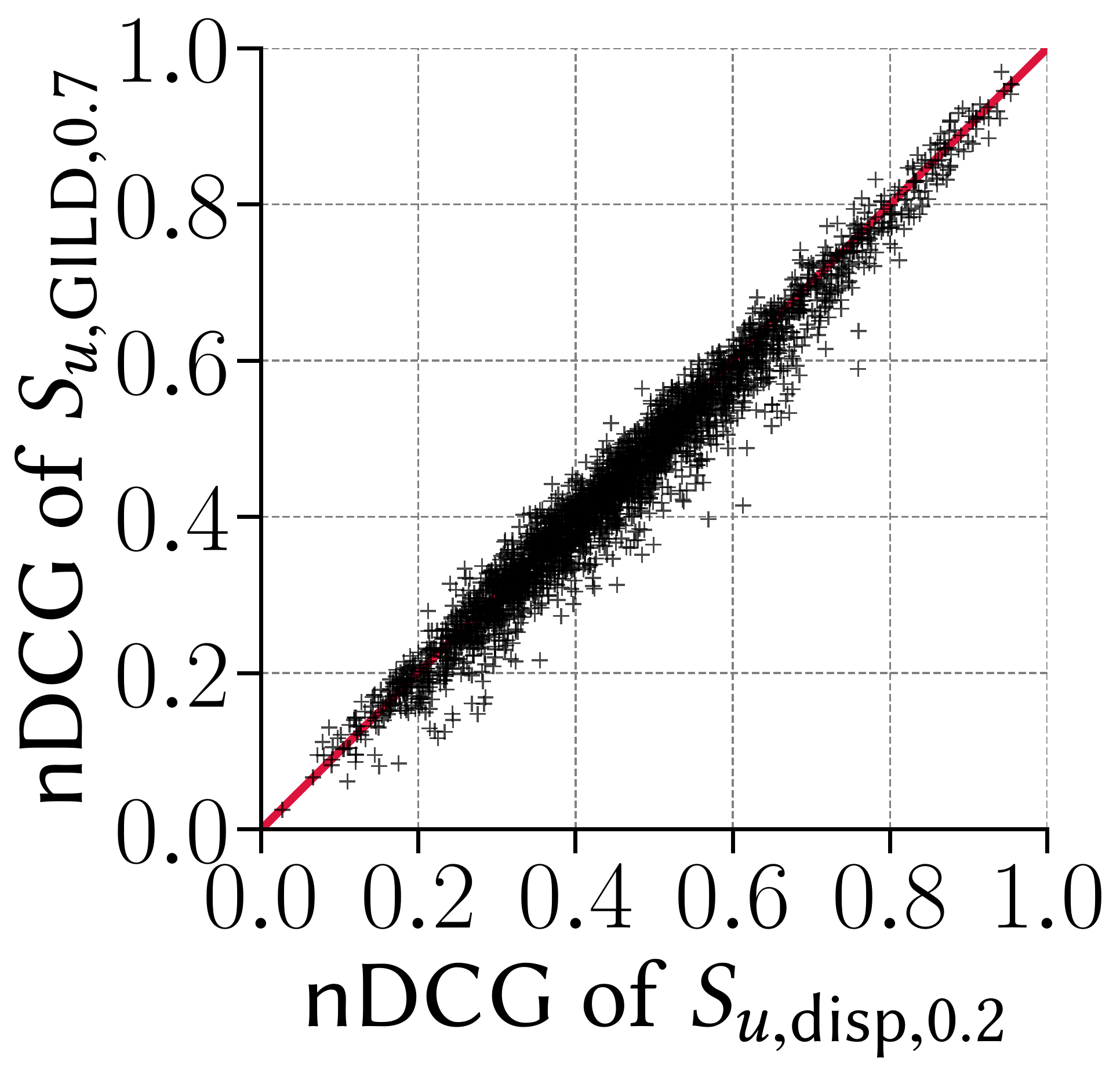}%
    \hfill%
    \includegraphics[width=.48\hsize]{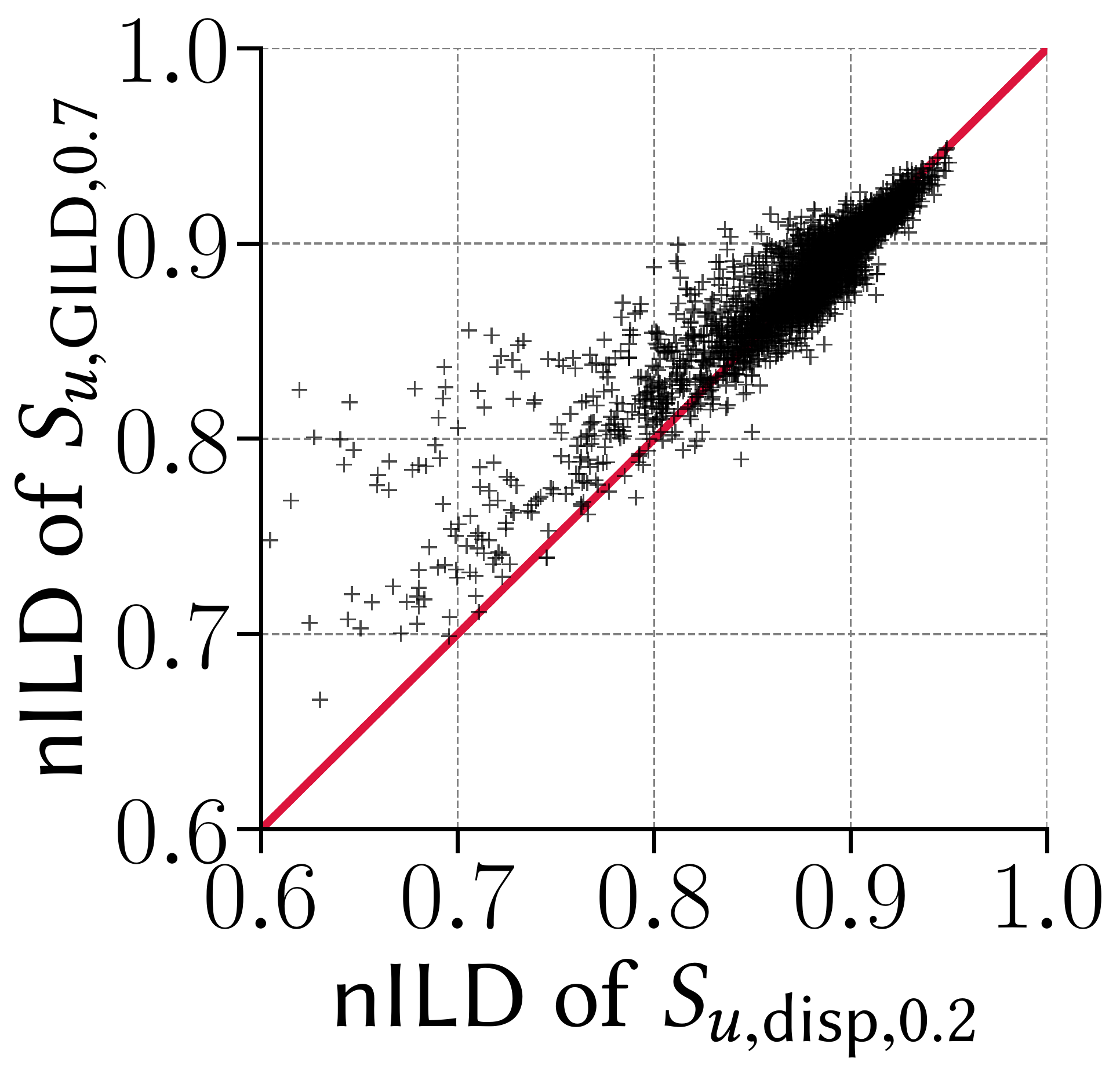}%
    \hfill\null
    \caption{
    Comparison of dispersion and GILD in terms of \nDCG and \nILD.
    }
    \label{fig:recommend_disp_vs_GILD}
\end{figure}

Although dispersion and GILD have a similar trade-off for the high-relevance case (i.e., mean $\nDCG \geq 0.4$), which is often a realistic situation,
they produce different results at the \emph{individual level}.
To this end, we select $\lambda$ such that
they are nearly identical \emph{on average}.
Specifically,
we choose $\lambda = 0.2$ for dispersion and $\lambda = 0.7$ for GILD, for which
the means of \nDCG, \nILD and \ndisp are respectively $0.457$, $0.870$ and $0.009$ for dispersion,
whereas
those are respectively $0.445$, $0.877$ and $0.001$ for GILD.
The left figure in \cref{fig:recommend_disp_vs_GILD} plots the \nDCG
of $S_{u,\disp,0.2}$ and $S_{u,\GILD,0.7}$ for each user $u$.
Observe that dispersion and GILD show a similar trend;
the standard deviation of \nDCG is $0.161$ for dispersion and $0.160$ for GILD.
In contrast, as shown in the right figure in \cref{fig:recommend_disp_vs_GILD},
dispersion often has a smaller \nILD than GILD.
Furthermore, the standard deviation of \nILD for dispersion ($0.051$)
is larger than that for GILD ($0.038$).
This difference is possibly due to the potential drawback of dispersion (see \cref{sec:theory:lessons}):
Since the values of dispersion for most users become $0$ at a particular iteration of the greedy heuristic,
the objective $\mathsf{F}_{u,\disp,0.2}(i)$ in \cref{eq:recommend:greedy} is $0.8 \mathsf{rel}_u(i)$ in the subsequent iterations; i.e.,
the greedy heuristic only selects the item with the highest relevance.
Consequently,
dispersion fails to diversify some users' recommendation results,
which is not the case for GILD.
In summary, as a diversity objective to be optimized in diversified recommendation,
ILD and dispersion are not an appropriate choice.

\section{Conclusions}
\label{sec:conclusions}
To investigate the behavior of two common diversity objectives, ILD and dispersion,
we performed a comparison analysis.
Our results revealed the drawbacks of the two:
ILD selects \emph{duplicate} items,
while dispersion may overlook \emph{distant} item pairs.
To analyze these drawbacks empirically,
we designed Gaussian ILD (GILD) as an interpolation between ILD and dispersion.
In the personalized recommendation setting,
we demonstrated that both ILD and dispersion are not 
consistently successful in enhancing diversity at the individual level.
As a future work,
we plan to develop an evaluation measure of diversity in lieu of ILD and dispersion.

\appendix
\section{Omitted Proofs in \NoCaseChange{\cref{sec:theory,sec:gauss}}}
\label{app:proof}

\begin{proof}[Proof of \cref{thm:ILD-disp-approx}]
The first guarantee is immediate from
$\OPT_{\ILD} \leq D$ and
$\ILD(S_{\disp}^*) \geq d_k^*$.
Similarly, we have
$
    \ILD(S_{\disp}^\Gr) \geq \disp(S_{\disp}^\Gr) \geq \frac{d_k^*}{2}
$
due to a $\frac{1}{2}$-approximation guarantee of the greedy heuristic \cite{ravi1994heuristic}.
Let $i_\ell \in S_{\disp}^\Gr$
denote the $\ell$-th item selected by greedy heuristic on \disp.
Since
$i_2$ is farthest from $i_1$,
$d(i_1, i_2) \geq \frac{D}{2}$.
By the triangle inequality of $d$, we have
$
d(i_1, i_\ell) + d(i_\ell, i_2)
\geq d(i_1, i_2)
$
for all $\ell \geq 3$.
Thus,
\begin{align*}
    \ILD(S_{\disp}^\Gr) & =
    {k \choose 2}^{-1}
    \left[
        d(i_1, i_2) + \sum_{3 \leq \ell \leq k} d(i_1, i_\ell) + d(i_\ell, i_2)
    \right] \\
    & \geq
    {k \choose 2}^{-1}\frac{D}{2}(k-1) = \frac{D}{k},
\end{align*}
implying that
$
    \frac{\ILD(S_{\disp}^\Gr)}{\OPT_{\ILD}} \geq \frac{1}{k}.
$
\end{proof}

\begin{proof}[Proof of \cref{clm:ILD-disp-tight}]
Let $n$ be a multiple of $4$ and $\epsilon > 0$ a small number.
Construct $2n$ vectors in $\bbR_+^{n+2}$, denoted
$\mat{X} \triangleq \{\vec{x}_1, \ldots, \vec{x}_{\frac{n}{2}}\}$ and $\mat{Y} \triangleq \{\vec{y}_1, \ldots, \vec{y}_{\frac{n}{2}}\}$,
each entry of which is defined as:
\begin{align*}
    x_i(j) \triangleq
    \begin{cases}
    \frac{\epsilon}{\sqrt{2}} & \text{if } j = i, \\
    \sqrt{\frac{1-\epsilon^2}{2}} & \text{if } j = n+1, \\
    0 & \text{otherwise,}
    \end{cases}
    \text{and }
    y_i(j) \triangleq
    \begin{cases}
    \frac{\epsilon}{\sqrt{2}} & \text{if } j = i+\frac{n}{2}, \\
    \sqrt{\frac{1-\epsilon^2}{2}} & \text{if } j = n+2, \\
    0 & \text{otherwise.}
    \end{cases}
\end{align*}
Observe that 
$\|\vec{x}_i - \vec{x}_j\| = \|\vec{y}_i - \vec{y}_j\| = \epsilon$ for all $i \neq j \in [\frac{n}{2}]$,
$\|\vec{x}_i - \vec{y}_j\| = 1$ for all $i, j \in [\frac{n}{2}]$,
and thus $D = 1$.
Consider selecting $k \triangleq \frac{n}{2}$ vectors from $\mat{X} \cup \mat{Y}$ so that ILD or dispersion is maximized.
Clearly,
$\OPT_{\ILD}$ is
${k \choose 2}^{-1} ( (\frac{k}{2})^2 + 2 {\nicefrac{k}{2} \choose 2} \epsilon ) = \Theta(1)$,
which is attained when we select $\frac{k}{2}$ vectors each from $\mat{X}$ and $\mat{Y}$.
By contrast, \emph{any} set of $k$ items has the same value of dispersion, i.e., $d_k^* \triangleq \epsilon$.
Hence, we may have $S_{\disp}^* = \{\vec{x}_1, \ldots \vec{x}_k\}$ in the worst case,
where $ \ILD(S_{\disp}^*) = \epsilon$.
Consequently,
it holds that
$\frac{\ILD(S_{\disp}^*)}{\OPT_{\ILD}} = \bigO(\epsilon) = \bigO\left(\frac{d_k^*}{D}\right)$.
When we run the greedy heuristic on dispersion,
we can assume that the first selected item is $\vec{x}_1$
without loss of generality.
Then, we would have selected $\vec{y}_i$ for some $i$ as the second item.
In the remaining iterations, we may select $k-2$ vectors all from $\mat{X}$ in the worst case,
resulting in
$
    \frac{\ILD(S_{\disp}^\Gr)}{\OPT_{\ILD}}
    = \frac{1}{\Theta(1)} {k \choose 2}^{-1} \Bigl((k-1) + {k-1 \choose 2}\epsilon\Bigr) = \bigO\Bigl(\frac{1}{k} + \frac{d_k^*}{D}\Bigr)
$.
\end{proof}

\begin{proof}[Proof of \cref{clm:disp-ILD-inapprox}]
Let $n$ be an even number at least $4$.
Construct $2n-2$ vectors in $\bbR_+$,
denoted
$\mat{X} = \{1, \ldots, (\nicefrac{n}{2} \text{ times}), \ldots, 1\}$,
$\mat{Y} = \{n, \ldots, (\nicefrac{n}{2} \text{ times}), \ldots, n\}$, and
$\mat{Z} = \{2, 3, \ldots, n-1\}$.
Selecting $k \triangleq n$ vectors from $\mat{X} \cup \mat{Y} \cup \mat{Z}$ so that the ILD value is maximized,
we have $S_{\ILD}^* = \mat{X} \cup \mat{Y}$.
Observe easily that the greedy heuristic selects at least two vectors from either  $\mat{X}$ or $\mat{Y}$.
Therefore, $\disp(S_{\ILD}^*) = \disp(S_{\ILD}^\Gr) = 0$.
By contrast, the optimum dispersion is $\OPT_{\disp} = 1$ and
attained when we select $\{1,2,\ldots,n\}$.
\end{proof}

\begin{proof}[Proof of \cref{thm:GILD}]
Let $S \triangleq [n]$.
We first calculate a limit of $\GILD_\sigma(S)$ as $\sigma \to \infty$.
Define
$
\epsilon_\sigma \triangleq \max_{i \neq j \in S} \frac{d(i,j)}{\sigma}.
$
Using a Taylor expansion of
$
    \exp\Bigl(-\frac{x^2}{2\sigma^2}\Bigr),
    =  1-\frac{x^2}{2\sigma^2} + \bigO\left( \frac{x^2}{\sigma^4} \right),
$
we derive
\begin{align*}
    \GILD_\sigma(S)
    & = {n \choose 2}^{-1} \sum_{i \neq j \in S}
    \sqrt{ \frac{d(i,j)^2}{\sigma^2} +\bigO\left( \frac{d(i,j)^4}{\sigma^4} \right) } \\
    & = \frac{\sqrt{1 + \bigO(\epsilon_\sigma^2)}}{\sigma} \cdot {n \choose 2}^{-1} \sum_{i \neq j \in S} d(i,j).
\end{align*}
Observing that 
$\lim_{\sigma \to \infty} \epsilon_\sigma = 0$,
we have
$
    \lim_{\sigma \to \infty} \frac{\GILD_\sigma(S)}{\frac{1}{\sigma} \ILD(S)} = 1,
$
completing the proof of the first statement.

We next calculate a limit of $\GILD_\sigma(S)$ as $\sigma \to 0$.
Define
$
    \delta \triangleq ( \min_{i \neq j \in S, d(i,j) > \disp(S)} \allowbreak d(i,j) ) - \disp(S).
$
Note that no pair of items $(i,j)$ satisfies $\disp(S) < d(i,j) < \disp(S)+\delta$.
Then define
$
    \epsilon_\sigma \triangleq \exp\left( -\frac{(\disp(S)+\delta)^2 - \disp(S)^2}{2 \sigma^2} \right).
$
Observe that for any pair $(i,j)$,
\begin{align*}
    \exp\left(-\frac{d(i,j)^2}{2 \sigma^2}\right) \text{ is }
    \begin{cases}
    =
    \exp\left(-\frac{\disp(S)^2}{2 \sigma^2}\right) & \text{if } d(i,j) = \disp(S), \\
    \leq 
    \epsilon_\sigma \exp\left(-\frac{\disp(S)^2}{2 \sigma^2}\right) & \text{otherwise.}
    \end{cases}
\end{align*}
Using a Taylor expansion of
$\sqrt{1+x} = 1 + \frac{1}{2}x \pm \bigO(x^2)$
yields
\begin{align}
\label{eq:GILD:expand}
\begin{aligned}
    & \GILD_\sigma(S)
    = \sqrt{2} - \frac{C}{2 \cdot {n \choose 2}} \cdot \exp\left(-\frac{\disp(S)^2}{2 \sigma^2}\right) \\
    & - \frac{{n \choose 2}-C}{2 \cdot {n \choose 2}} \cdot \bigO\left(\epsilon_\sigma \cdot \exp\left(-\frac{\disp(S)^2}{2\sigma^2}\right)\right)
    \pm \bigO\left( \exp\left(-\frac{\disp(S)^2}{2\sigma^2}\right)^2\right),
\end{aligned}
\end{align}
where $C$ is the number of pairs $(i,j)$ with $d(i,j) = \disp(S)$.
Observing that
$\lim_{\sigma \to 0} \epsilon_\sigma = 0$,
we have
\begin{align*}
    \lim_{\sigma \to 0} \frac{\GILD_\sigma(S) - \sqrt{2}}{ - \frac{C}{2 \cdot {n \choose 2}} \cdot \exp\left(-\frac{\disp(S)^2}{2 \sigma^2}\right) } = 1,
\end{align*}
completing the proof of the second statement.
\end{proof}

\onecolumn
\begin{multicols}{2}
\bibliographystyle{ACM-Reference-Format}
\bibliography{main}
\end{multicols}

\end{document}